\documentclass[pdflatex,preprint,flushrt]{aastex} 
\usepackage{natbib} 
\usepackage{epsfig} 
\usepackage{float} 

\usepackage{graphicx}
\usepackage{graphics}
\usepackage{subfigure}
\usepackage{rotating} 
\usepackage{empheq}

\usepackage{verbatim}
\newcommand{\fermi}{{\it Fermi} Gamma-ray Space Telescope} 
 
\newcommand{\latf}{{\it Fermi}-LAT} 
\newcommand{\sax}{{\it BeppoSAX}} 
\newcommand{\swf}{{\it Swift}} 
\newcommand{\swx}{{\it Swift--XRT}} 
\newcommand{\swu}{{\it Swift--UVOT}} 
\newcommand{\swb}{{\it Swift--BAT}}

\newcommand{\asca}{{\it ASCA}} 
\newcommand{\suzaku}{{\it Suzaku}} 
 
\newcommand{\EminLat}{200 MeV} 
\newcommand{\ST}{v9r15p5} 
\newcommand{\IRFs}{P6\_V3\_DIFFUSE} 
\newcommand{\gtlike}{\texttt{gtlike}} 
\newcommand{\PKS}{PKS~1510-089} 
\newcommand{\PKSt}{PKS~1510-089~} 
\def\g-ray{$\gamma-$ray} 

\def\latflux{ph cm$^{-2}$ s$^{-1}$}
\def\flux{erg cm$^{-2}$ s$^{-1}$}
\def\lum{erg s$^{-1}$}

\def\latfluxvii{$\times 10^{-7}$ ph cm$^{-2}$ s$^{-1}$}

\title{{\it Fermi} Large Area Telescope and multi-wavelength observations of the flaring activity of \PKS~between 2008 September  and 2009 June } 

\author{
A.~A.~Abdo\altaffilmark{2,3}, 
M.~Ackermann\altaffilmark{4}, 
I.~Agudo\altaffilmark{5,51}, 
M.~Ajello\altaffilmark{4}, 
A.~Allafort\altaffilmark{4}, 
H.~D.~Aller\altaffilmark{6}, 
M.~F.~Aller\altaffilmark{6}, 
E.~Antolini\altaffilmark{7,8}, 
A.~A.~Arkharov\altaffilmark{9}, 
M.~Axelsson\altaffilmark{10,11,12}
% ...
}
% ...
\altaffiltext{2}{Space Science Division, Naval Research Laboratory, Washington, DC 20375, USA}
\altaffiltext{3}{National Research Council Research Associate, National Academy of Sciences, Washington, DC 20001, USA}
\altaffiltext{4}{W. W. Hansen Experimental Physics Laboratory, Kavli Institute for Particle Astrophysics and Cosmology, Department of Physics and SLAC National Accelerator Laboratory, Stanford University, Stanford, CA 94305, USA}
\altaffiltext{5}{Instituto de Astrof\'{\i}sica de Andaluc\'{\i}a, CSIC, Apdo. 3004, Granada 18080, Spain}
\altaffiltext{6}{Department of Astronomy, University of Michigan, Ann Arbor, MI 48109-1042, USA}
\altaffiltext{7}{Istituto Nazionale di Fisica Nucleare, Sezione di Perugia, I-06123 Perugia, Italy}
\altaffiltext{8}{Dipartimento di Fisica, Universit\`a degli Studi di Perugia, I-06123 Perugia, Italy}
\altaffiltext{9}{Pulkovo Observatory, 196140 St. Petersburg, Russia}
\altaffiltext{10}{Department of Astronomy, Stockholm University, SE-106 91 Stockholm, Sweden}
\altaffiltext{11}{Lund Observatory, SE-221 00 Lund, Sweden}
\altaffiltext{12}{The Oskar Klein Centre for Cosmoparticle Physics, AlbaNova, SE-106 91 Stockholm, Sweden}
% ...
\altaffiltext{51}{Institute for Astrophysical Research, Boston University, Boston, MA 02215, USA}
% ...

\begin{abstract}
We report on   the multi-wavelength observations of   \PKS~ (a flat spectrum radio quasar at $z$=0.361) 
during its high activity period between 2008 September and 2009 June. During this  11 months period,
the source was characterized  by a complex variability  at optical, UV and \g-ray bands, on 
time scales  down to 6-12 hours. 
The brightest \g-ray isotropic luminosity, recorded on 2009 March 26, 
was $\simeq 2\times 10^{48}$\lum. 
The spectrum in the {\it Fermi}-LAT energy range shows a mild curvature well described
by a log-parabolic law, and can be understood as due to the Klein-Nishina effect.
The \g-ray flux has a complex correlation with the other
wavelengths. There is no correlation at all with the X-ray band,
a weak correlation with the UV, and a significant  correlation with the
optical flux. The \g-ray flux seems to lead the optical 
one by about 13 days. 
From the UV photometry we estimated a black hole mass of 
 $\simeq 5.4\times 10^8 M_{\odot}$, and
an accretion rate of $\simeq 0.5~ M_{\odot}$/yr.
Although the power in the thermal and non-thermal outputs is smaller 
compared to the very luminous and distant flat spectrum radio quasars, 
\PKS~exhibits a quite large Compton dominance and a prominent big blue bump (BBB) 
as observed in the most powerful \g-ray quasars.
The BBB was still prominent during the historical maximum optical state in 2009 
May, but the optical/UV spectral index was softer than in the quiescent state. 
This seems to indicate that the BBB was not completely dominated
by the synchrotron emission during the highest optical state. 
We model the broadband spectrum assuming a leptonic scenario in
which the inverse Compton emission is dominated  by the scattering 
of soft photons produced externally to the jet.
The resulting model-dependent jet energetic content is compatible with a scenario in which
the jet is powered by the accretion disk, with a total efficiency 
within the Kerr black hole limit.

\end{abstract} 

\begin{document}
\section{Introduction} \label{sec:Intro} 
Among blazars, flat spectrum radio quasars (FSRQ) are those objects characterized 
by prominent emission lines in the optical spectra. The typical spectral energy distribution 
(SED) of blazars has a two bump shape. According to current models the low energy bump is 
interpreted as synchrotron  emission from highly relativistic electrons, and the high 
energy bump is interpreted as inverse Compton (IC) emission. 
In FSRQs the IC bump can dominate over the synchrotron one by more 
than an order of magnitude. It is widely believed that 
in these sources the IC component is dominated by the scattering of soft photons produced 
externally to the jet \citep{Sikora1994,DermerCharlesD.2002}, 
rather than by the synchrotron self Compton emission \citep[SSC,][]{Jones1974,Ghisellini1989}.
In the external radiation Compton scenario (ERC), the seed photons for the IC process 
are typically UV photons generated by the accretion disk surrounding the black 
hole, and reflected toward the jet by the broad line region (BLR) within a 
typical distance from the disk in the sub pc scale. If the 
emission occurs at larger distances, the external radiation is likely to be 
provided by a dusty torus (DT) \citep{SikoraM.2002}. In this case the radiation 
is typically peaked at IR frequencies.

The study of the SEDs of blazars and their 
complex variability has been greatly enriched since the 2008 August start 
of scientific observations by the Large Area Telescope \citep[LAT,][]{Atwood2009} 
on the \fermi~ \citep{Ritz2007}, thanks to its high sensitivity and survey mode.
 
One of the most active blazars observed in this period was the FSRQ \PKS.  
This object has an optical spectrum characterized by prominent emission lines  
overlying a blue continuum \citep{TadhunterC.N.1993}  at a  redshift $z = 0.361$ \citep{Thompson1990}.
Radio images show a bright core with a jet which has a large misalignment between the arcsecond and
milliarcsecond scales; superluminal velocity up to $\simeq 20c$ are also reported \citep{Homan2002}.

\PKS~was already detected in \g-rays by EGRET \citep{Hartman1999} and 
exhibited a very interesting activity at all wavelengths. It was also detected by 
AGILE during ten days of pointed observations from 2007-08-23 to 2007-09-01 \citep{Pucella2008a}. 
In the period 2008-2009 \PKS~was observed to be bright 
and highly variable in several frequency bands. In gamma rays it was 	 
detected in 2008 March  by AGILE \citep{D'AmmandoF.2008} and other 
bright phases were observed in the subsequent months by both \textit{Fermi}-LAT and AGILE 
\citep{TramacereAndrea2008,CipriniS.2009, D'AmmandoF.2009, 
PucellaG.2009,VercelloneS.2009, CutiniS.2009}. High states in X-rays and 
in optical were reported by \cite{KrimmH.A.2009}, \cite{VillataM.2009}, and 
\cite{LarionovV.M.2009,LarionovV.M.2009a}. In a recent paper, \cite{Marscher2010} presented data from a multi-wavelength
(MW) campaign concerning the same flaring period  of \PKS. In that paper, the authors focus on analysis of the parsec scale 
behavior and correlation of rotation of the optical polarization angle 
with the dramatic \g-ray activity.

In the present paper we describe the results of the LAT monitoring together 
with the related multi-wavelength campaigns covering the entire  electromagnetic 
spectrum.  We present a detailed analysis of the $\gamma$-ray spectral shape and spectral
evolution, and of the MW SED modelling and interpretation. This paper is organized as follows:
in Section 2, we report results on the $\gamma$-ray observation of 
\PKS, and we study the spectral shape and its evolution. 
In Section 3, we summarize multifrequency 
data obtained through simultaneous optical-UV-X-ray \swf~ observations, and 
radio-optical observatories. In Section 4 we present the results of the 
multifrequency data and their connection with
the \g-ray activity. In Section 5 we report our conclusions about the
MW data, and we use a phenomenological analysis to estimate some of the physical fundamental
parameters, such as the black hole mass, the accretion disk bolometric
luminosity, the shape of the electron distribution and the beaming factor. 
We then model the observed SEDs, and we comment on the jet energetics. 
Furthermore, we compare \PKSt with other
powerful FSRQs observed by \textit{Fermi}. In Section 6 our final 
remarks are reported. 

In the following we use a $\Lambda$CDM (concordance) cosmology with values 
given within 1 $\sigma$ of the WMAP results \citep{Komatsu2009}, namely $h$ = 
0.71, $\Omega_m$ = 0.27, and $\Omega_{\Lambda}$ = 0.73, and a Hubble constant 
value $H_0$=100$h$ km s$^{-1}$ Mpc$^{-1}$, the corresponding luminosity distance 
($d_L$) is $\simeq 1.91$ Gpc ($\simeq 5.9\times 10^{27}$ cm).

\section{{\it Fermi}-LAT data and results} \label{sec:Fermi} The LAT data   
presented here were collected from 2008-08-04 to 2009-07-01. Only events with 
energies greater than \EminLat~were selected to minimize the systematics 
uncertainties. To have the highest probability that collected events are photons the 
{\it diffuse} class selection was applied. A further selection on the zenith 
angle $>105^{\circ}$ was applied to avoid contamination from limb 
gamma rays. The analysis was performed using the Science Tools  package\footnote{http://fermi.gsfc.nasa.gov/ssc/data/analysis/documentation/Cicerone/} (\ST). 
The Instrument Response Functions (IRF) \IRFs~were used. These IRFs provide a 
correction for the pile-up effect. To produce light curves and spectral 
analysis the standard tool \gtlike~was used. The photons were extracted from a 
region of interest (ROI) centered on the source, within a radius of 
$7^{\circ}$. The \gtlike~model includes the \PKS~point source component, and 
all the point sources form the first LAT catalog \citep{1LATCAT} that fall within
$12^{\circ}$ from the source.
The model includes also a background component of the Galactic diffuse emission and an 
isotropic component, both of which are the standard models available from the \textit{Fermi} Science
Support Center\footnote{http://fermi.gsfc.nasa.gov/ssc/data/access/lat/BackgroundModels.html} (FSSC). 
The isotropic component includes both the contribution 
from the extragalactic diffuse emission and from the residual charged particle 
backgrounds. The estimated systematic uncertainty of the flux is $10\%$ at 
100 MeV $5\%$ at 500 MeV and $20\%$ at 10 GeV. 

\subsection{Temporal behavior}
\label{sec:Fermi-time}
We extracted light curves from the entire data set, to investigate the flaring 
activity. To take into account possible biases or systematics when the source 
flux is faint, we used two different time binnings of one day and one week. 
The light curves were extracted using \gtlike~, fitting the 
source spectrum by means of a power-law distribution ($dN/dE \propto 
E^{-\alpha_{\gamma}}$), {where $\alpha_{\gamma}$ is the photon index}, 
following the prescription given in the previous Section. The 
flux was evaluated by integrating the fitted model above 0.2 GeV. The lower 
panel of Fig. \ref{fig:LATLC} shows clearly four major flaring episodes: 
between 2008-08-30 and 2008-09-26 (flare $a$), between 2009-01-04 and 
2009-01-27 (flare $b$), between 2009-03-10 and 2009-04-09 (flare $c$), and 
between 2009-04-15 and 2009-05-12 (flare $d$). The source was almost quiescent 
between the end of 2008 Sep. and the beginning of 2009 Jan. (see. Tab. 
\ref{tab:flares} for a summary). { The flux light curves with different temporal 
binning are compatible, with the daily integration binning showing better rapid 
flux variations that are smoothed in the weekly binning}. 
A study of these variations based on the autocorrelation, Fourier analysis 
and structure function is presented by \cite{Abdo2010a} together 
with other blazars.

\begin{figure}[H]
\begin{center}
\epsfig{file=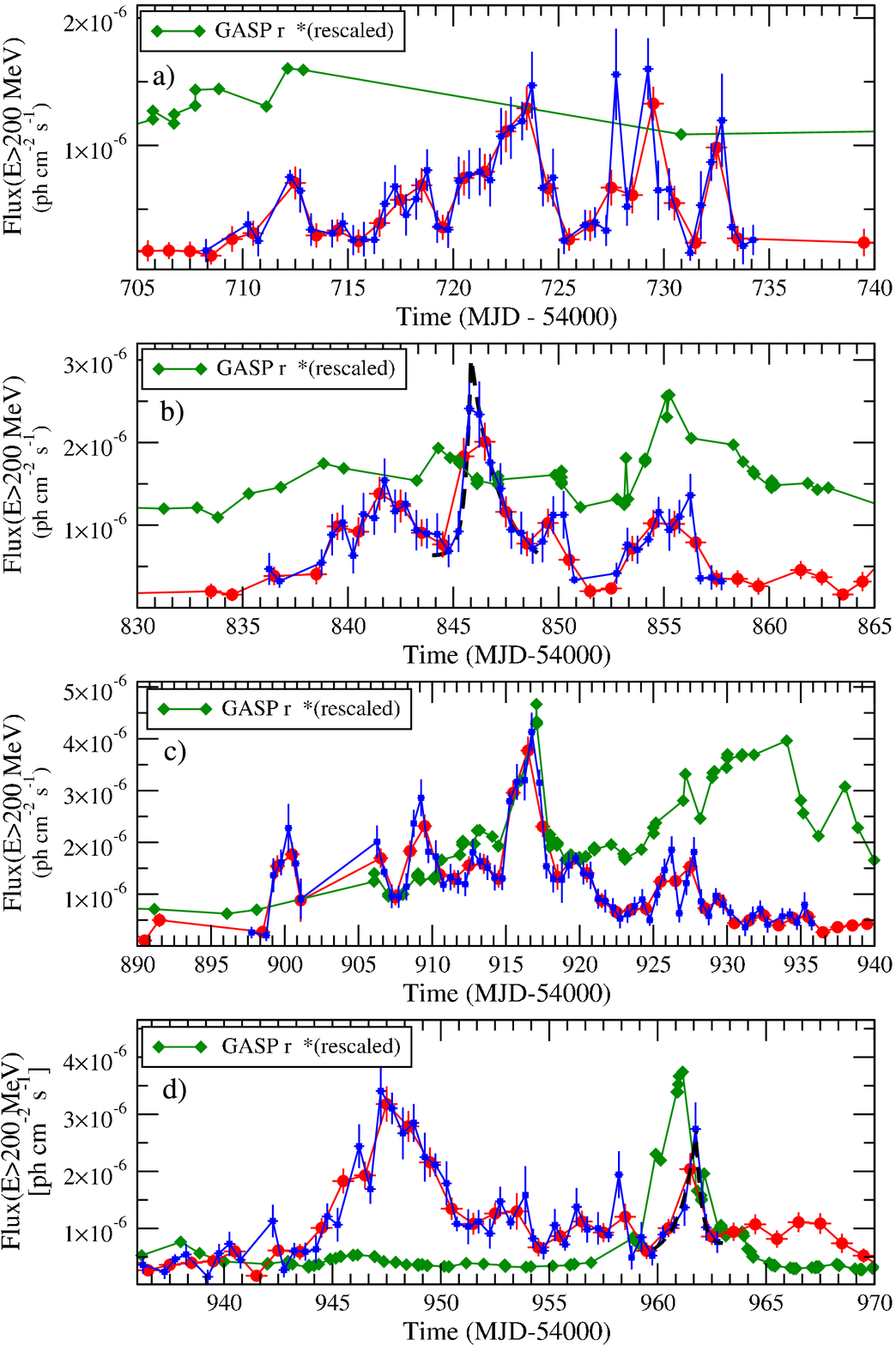,width=15cm,angle=-0}
\end{center}
\caption{from top to bottom, flare \textit{a, b, c,} and \textit{d} showing 1-day binning (red points) 
and 12 hr binning (blue points). The green points represent the optical data in the \textit{R}
filter. The black dashed lines represent a best-fit by means of an exponential law as described 
in Sect. \ref{sec:Fermi-time} }
\label{fig:LAT_SING_FLARES} \end{figure}

Fig. \ref{fig:LAT_SING_FLARES} shows close-up light curves of the four flares 
and the green line represents the optical data in the \textit{R} filter (see Sect. 
\ref{sec:MWDATA} and \ref{sec:MWRESULTS}). Since the statistics during the 
flares were high, it was possible to use also a 12 hr binning (blue points). 
Typically, the flares have a complex structure with peaks having durations from 
about 1 to 5 days and their moderately asymmetric profile can result from the 
overlapping of subsequent episodes. In a few cases, significant variations by a 
factor of 2 within 12 hr were detected. To have a better estimate of the 
rising and decaying timescales, we fit two rapid flares, with an almost 
regular shape, using an analytical law of the form $A \cdot\exp^{t/\tau}$. 
In the case of the flare peaking at $t\simeq$ 54846 MJD (dashed black line in panel $b$ 
of figure \ref{fig:LAT_SING_FLARES}) we find a rise faster than the decay: 
the rising timescale is $\tau \simeq 0.3$ 
days and the decaying one is $\simeq 1.4$ days. The second event 
(flare peaking at $t\simeq$ 54962 MJD, dashed black line in the $d$ panel of 
\ref{fig:LAT_SING_FLARES}) followed the 
opposite behavior, having a best fit rising timescale of $\simeq 0.8$ days,
and a decay time of $\simeq 0.25$ days. {Using a bin width of 6 hr, the shape 
of the flare is nearly symmetric, with rise and decay \textit{e}-folding time 
of about $\simeq 0.12$ days.} 
Such a fast variability can constrain the radiative region size $R_{rad}$ 
by the well known relation 

\begin{equation} 
\label{eq:size_constr} R_{rad} \leq 
\frac{c~\Delta t~\delta}{1+z} , 
\end{equation} 

\noindent where $c$ is the speed of the 
light, $\delta=1/(\Gamma(1-\beta\cos \theta))$ is the beaming factor depending
on the bulk Lorentz factor $\Gamma$ and a viewing angle $\theta\simeq 1/\Gamma$, 
 and $z$ is the cosmological redshift. Adopting the very fast superluminal 
velocity reported by \citet{Homan2002} ($\beta_{app}\simeq 20$, in very good agreement 
with the results presented in Section \ref{sec:mm-Radio results}) from 
$\delta\simeq\Gamma\simeq\beta_{app}$ we can estimate $R_{rad} \lesssim 9\times 
10^{15}$ cm (in the case of $\tau \simeq 0.25$ days). 
We will compare this result with other constraints derived in Sect. \ref{sec:gamma-ray-phys}.

\subsection{Gamma-ray Spectra}
\label{sec:Fermi-Gamma-ray-spectra} 
\begin{figure}[H]
\begin{center}
\epsfig{file=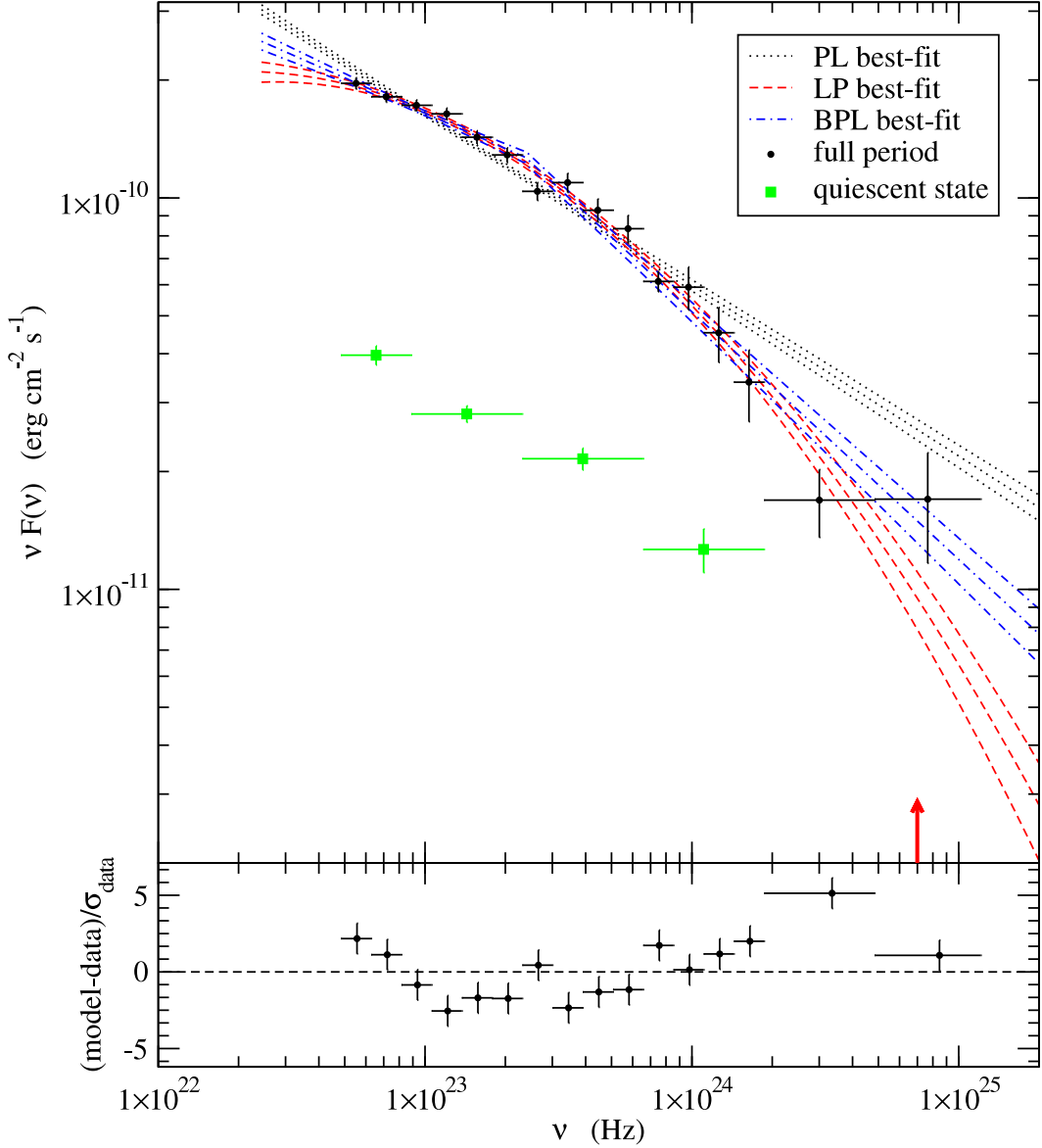,angle=-0,width=14cm}
\end{center}
\caption{LAT SED of \PKS~extracted for the full period (black points) and for 
the quiescent state (green points). The red upward arrow indicates the 
highest energy event within 95$\%$ of the PSF for the whole period data set. 
The dotted line, the dashed line, and the dot-dashed line, represent the best 
fit model of the full period by means of PL, LP, a BPL distribution 
respectively, with uncertainties. The residuals in the lower panel refers to the PL model.} 
\label{fig:LAT_SED} \end{figure}

\begin{figure}[H]
\begin{center}
\epsfig{file=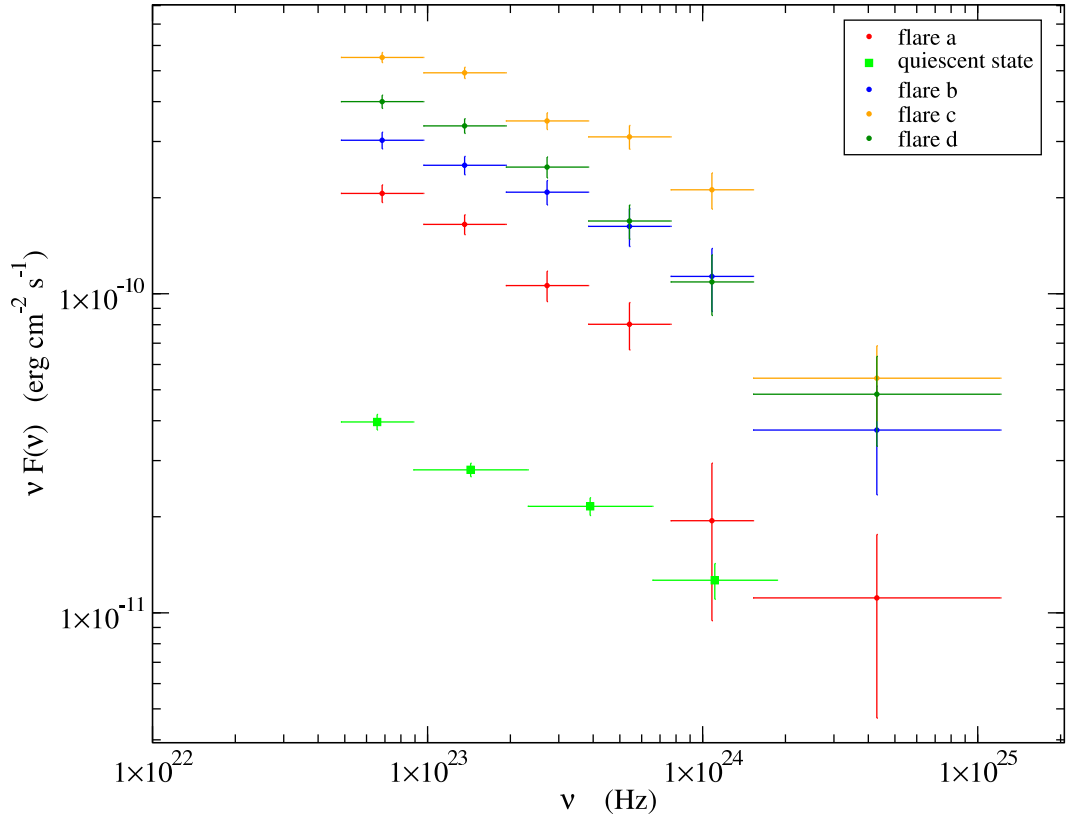,angle=0,width=14cm}
\end{center}
\caption{ LAT SED of \PKS~extracted for the flaring states and for the 
quiescent state (green points).} 
\label{fig:LAT_SED_FLARES} 
\end{figure}

We analysed the $\gamma$-ray spectral shape of \PKS~during the whole period, the 
quiescent state, and the four flaring episodes using three spectral models: a 
power-law (PL) a log-parabola (LP), $dN/dE \propto E/E_0^{-\alpha_{\gamma}-\beta 
\log(E/E_0)}$ \citep{Landau1986,Massaro2004}, and a broken power-law (BPL).
In the case of LP spectral law, the parameter 
$\beta$ measures the curvature around the peak. 
The LP distribution has only three free parameters, and the 
choice of the reference energy $E_0$ does not affect the spectral shape; we 
fixed its value to 300 MeV.  {We performed the spectral analysis using an 
unbinned maximum-likelihood estimator (\gtlike) and the same 
prescription given in Sec. \ref{sec:Fermi}. We used a likelihood ratio test
\footnote{The LRT statistic is defined as: LRT = -2Log($L0$/$L1$), where $L0$ and $L1$ are the maximum likelihood 
estimated for the null and alternative hypothesis respectively.} 
\citep[LRT,][]{mattox1996} to check the PL model (null hypothesis) against the LP 
model (alternative hypothesis). Since the PL is often
rejected, we test also the LP model (null hypothesis) against the BPL model (alternative hypothesis).}
The results concerning the LRT are summarized in Tab. \ref{tab:LAT_LRT}, and in Tab. \ref{tab:LAT_FIT} 
we report the details of the spectral analysis for each time range and 
spectral model. Due to the non derivable character of the BPL 
law, we used also the loglikelihood profile method to determine the best fit 
parameter for this model. The corresponding statistical uncertainty was 
estimated from the difference in the likelihood value w.r.t. its minimum such 
that $-2\Delta L$=1.  The $\gamma$-ray spectrum of \PKSt is well described by a 
log-parabola, with the only exception being the quiescent state. The value of the 
LRT reported in Tab. \ref{tab:LAT_LRT} shows that both for the flares ($a$, 
$b$, $c$, $d$) and for the full period, the LP model describes the spectrum 
better than PL with a probability higher than $\simeq 99.6\%$. The BPL, on the 
contrary, does not provide an improvement w.r.t. the LP model. The only 
exception is the flare $b$, but the false positive probability is about \textit{$27.33\%$}, 
so there is no evidence to reject the null hypothesis.
The LP model is then preferred because of the lower number of parameters.
Moreover the curvature parameter $\beta$ can be linked to physical 
processes such as the acceleration or the effects of the Klein-Nishina regime in the IC process, 
as we will discuss in Sect. \ref{sec:gamma-ray-phys} and \ref{sec:SED-Model}.  
For a better visualization of the SED shape and to show the 
departure from a PL trend, we produced an SED by performing an independent 
likelihood analysis starting form a grid of 20 energy bins logarithmically equispaced. 
The bins were then grouped in order to have at least 10 photons per bin, and the highest energy bin was chosen 
according to the maximum energy encircled within 95$\%$ of the PSF. The 
results are shown in Fig. \ref{fig:LAT_SED} and Fig \ref{fig:LAT_SED_FLARES}.  
In Fig. \ref{fig:LAT_SED} we show the full-period SED and the one extracted 
during the quiescent state. We plot by a dotted line the PL model, by a dashed 
line the LP model, and by a dot-dashed line the BPL model. With a red upward 
arrow we indicate the highest energy event within $95\%$ of the PSF for the 
whole period data set, corresponding to an energy of approximately 30 GeV. From the 
plot of the PL model residuals (lower panel) it is possible to see clearly the 
departure from a PL trend: the deviations, both at low and high energies, 
suggests for a spectral curvature confirming the LRT results. Moreover, it is 
possible to note that the BPL model does not deviate significantly from 
the LP trend, supporting again the LRT analysis. The SEDs of individual flares are 
plotted in Fig. \ref{fig:LAT_SED_FLARES} and show that the spectrum was curved 
also during the single flaring episodes. We note that the flare-integrated 
spectral shape did not change significantly, despite the huge flux variations. 

\subsubsection{Spectral evolution.}
\label{sec:Fermi-Spec-Ev} 
\begin{figure}[H]
\begin{center}
\epsfig{file=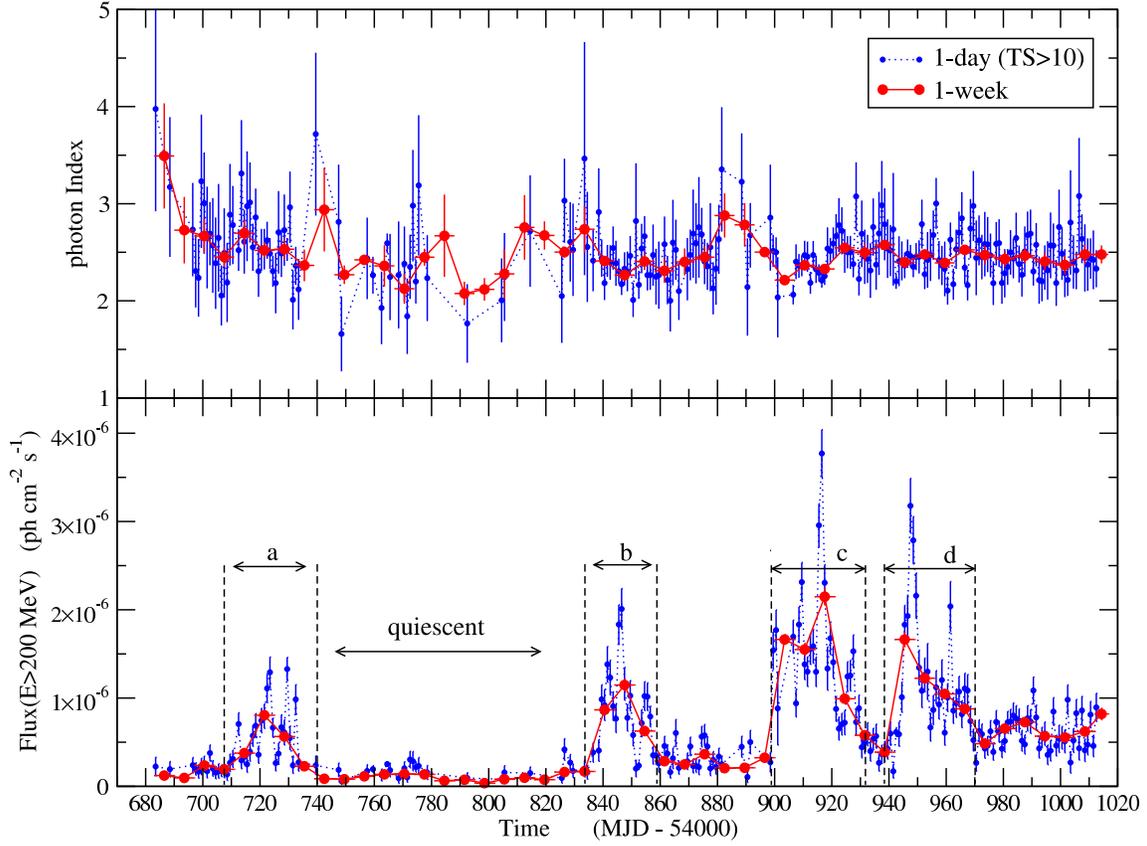,width=15cm,angle=0}
\end{center}
\caption{{\it Upper Panel}:{ the \g-ray photon index ($\alpha_{\gamma}$), as a function of the time, 
for weekly and daily binning. In case of daily binning only observations with a Test Statistic $>10$ are
taken into account. The Test Statistic is defined as TS: = -2 Log($L0/L1$), where $L1$ and $L0$ are the likelihood when the source is included or not}. {\it Lower Panel}: weekly, and 
daily( TS$>$10) fluxes light curve.} 
\label{fig:LATLC} 
\end{figure}

\begin{figure}[H]
\begin{center}
\epsfig{file=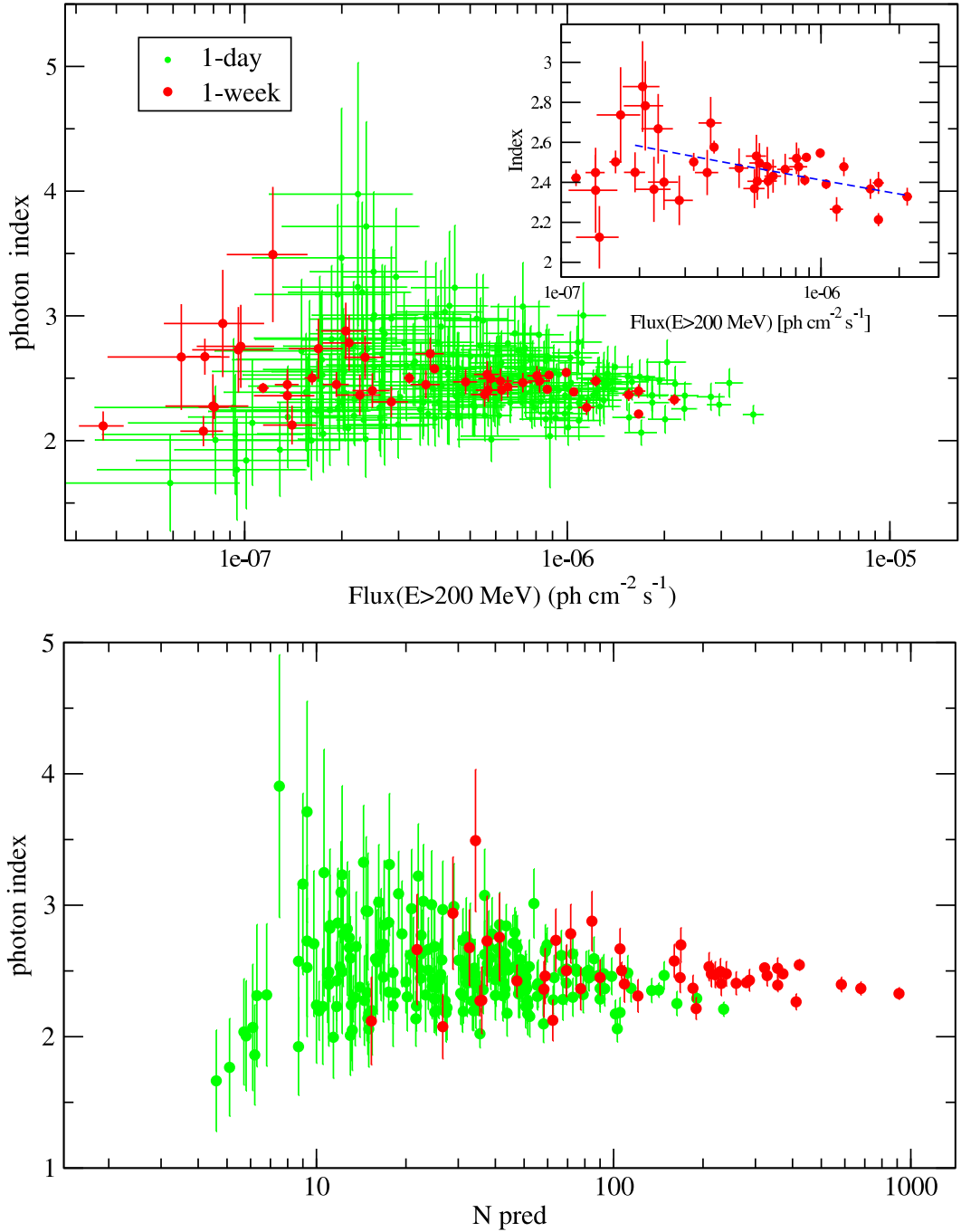,width=14cm}
\end{center}
\caption{{\it Upper Panel}: weekly and daily scatter plot of the Flux( 
E$>200$ MeV) vs. the photon index (TS $>10$). The inset shows the
same for Flux( E$>200$ MeV) $>$ 2\latfluxvii . {\it Lower Panel}: 
Scatter plot of the number 
of photons predicted by the best-fit model vs. the photon index, for weekly and 
daily integration (TS $>$10).
} 
\label{fig:LAT_INDEX_VS_FLUX_N} 
\end{figure}

To complete the analysis of the spectral behavior of \PKS, we investigate 
whether spectral changes are seen between the quiescent and the flaring 
state. Since we are mainly interested in the search of possible trends rather 
than in the best description of the spectral distribution, we simply evaluated 
the PL spectral indices in the various brightness states, which can be 
considered representative of the mean slope. In the upper panel of Fig. 
\ref{fig:LAT_INDEX_VS_FLUX_N} we plot the photon index against the flux 
above 0.2 GeV, resulting from the same spectral analysis showed in Fig. 
\ref{fig:LATLC}. In the case of daily integration (green circles), and
more marginally for the weekly integration (red circles), this plot 
seems to show a softer when brighter trend, up to a flux level of $F(E>0.2 GeV) 
\simeq$ 2.4\latfluxvii.
Above this value of the flux the source
has a harder when brighter trend. We analyse the correlation of the
harder when brighter trend, for the weekly binning, using a Monte Carlo method 
that takes into account the dispersion of flux and
index measurements. In detail, we re-sample the flux and index
values for each observed pair, extracting the data
from a normal distribution 	centered on the observed value,
and with the standard deviation equal to the 1 $\sigma$ error
estimate.
We find a correlation coefficient of $r=0.43$
with a $95\%$ confidence limit 0.24 $\leq r \leq$ 0.58. 
The trend for $F(E>0.2 GeV) \gtrsim $ 2.4\latfluxvii is reported in the inset 
of the upper panel of Fig. \ref{fig:LAT_INDEX_VS_FLUX_N}.\\
Although for some EGRET blazars \cite{Nandikotkur2007} observed a similar 
flux-hardness anticorrelation at low fluxes,
we need to take into account possible effects coming from the poor statistics 
when the source flux is low. As first, we note that moving to the weekly binning 
the trend is less evident, although the flux range is the same as that of the
daily binning. As a further check, in the lower panel of the Fig. \ref{fig:LAT_INDEX_VS_FLUX_N} 
we plot the photon index against the number of photons predicted 
by the best-fit model. It is clear that the dispersion of the photon index 
is related closely to the number of predicted photons, 
above $N\simeq 20$ the trend is the same for both the 2 integration timescales, 
and the photon index clusters around 2.5, without showing the 
very soft and very hard indices values present at low number of events. 

In conclusion, we can not exclude the presence of a softer when brighter trend, for 
low flux levels and short timescales, {but the statistical effects we present 
do not allow to obtain a purely physical interpretation.} 
Similar results have been found for other \textit{Fermi} Blazars \citep{Abdo2010}, independently 
of their redshift or class (BL Lacs/FSRQs). 

\begin{figure}[H]
\begin{center}
\epsfig{file=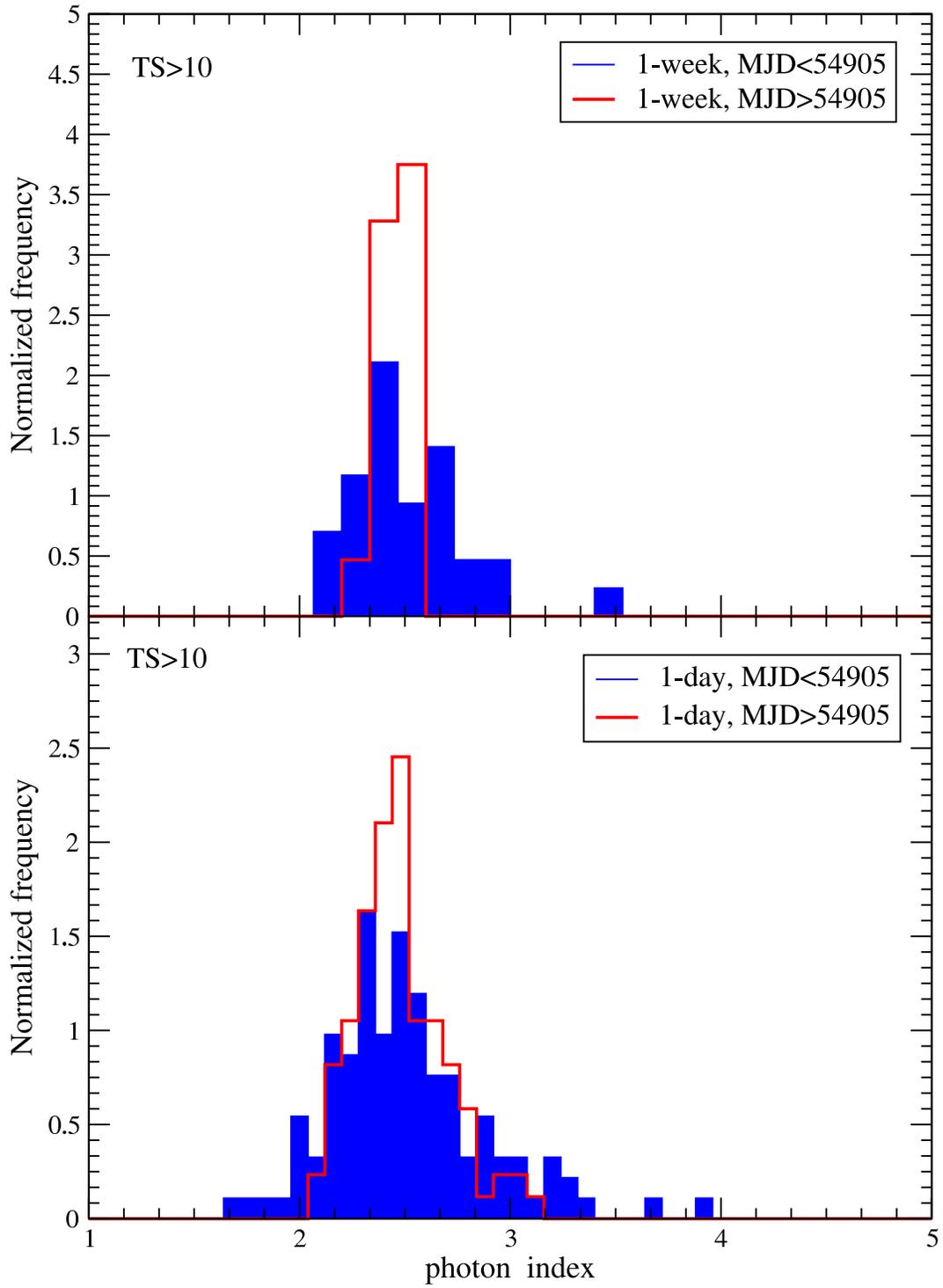,width=14cm,angle=0}
\end{center}
\caption{{\it Upper Panel}: The histogram of the photon index for,
a weekly integration time, before (blue) and after (red) MJD
54905, respectively. {\it Upper Panel}: The same as in the upper panel, in the
case of daily integration.}
\label{fig:LAT_INDEX_HISTO}
\end{figure}

Even if we did not find a strong evidence for a index-flux correlation, 
the variation of the photon index returned by the \gtlike~ fit as a function of 
time (see upper panel of Fig \ref{fig:LATLC}), shows that the dispersion 
on the photon index is larger until mid March roughly, and gets narrower 
after. This feature is emphasized in Fig. 
\ref{fig:LAT_INDEX_HISTO}, where in the upper panel we plot the histogram of 
the photon index for the weekly integration, before MJD 54905 (corresponding 
to 2009-03-15, blue shaded histogram), and after MJD 54905 (red empty 
histogram). In the lower panel we plot the same analysis for the case of daily 
integration.  The distributions before and after MJD 54905 have the same mean, 
in both daily and weekly integrations ($\simeq 2.5$), but very different standard 
deviations. In the case of daily integration we have 0.45 and 0.21 before and 
after MJD 54905, respectively. In the case of the weekly integration we have a 
standard deviation of 0.28 before MJD 54905 and 0.07 after.  
To test more quantitatively whether or not the distributions are different, 
we use a Kolmogorov-Smirnov (KS) test. We applied { the test} to the distributions of the 
spectral indices before and after MJD 54905, for the daily and 
weekly integrations. The test returns a p-value of $\simeq 0.13$ and 
$\simeq 0.35$, for the case of weekly and daily binning respectively. 
The KS test gives only a marginal indication that the data sets, before and after MJD 54905, 
may  not be drawn from  the same distribution. 

In conclusion, the typical $\gamma$-ray photon index of \PKS~is around 2.5
and values largely different from this were never observed in high states.
LP best fits indicate a significant but quite mild spectral curvature, that 
in the brightest flares ($b$, $c$ and $d$) was always very close to $\beta=$ 0.1.

\section{Multifrequency observations and data reduction} 
\label{sec:MWDATA} 
The unique, high quality data provided by the LAT 
instrument can not be physically fully understood without simultaneous 
multifrequency observations. The spectral curvature and the spectral evolution 
observed in the $\gamma$-ray band, need to be compared to SED evolution from the radio 
to the hard X-ray. Radio data, and in particular VLBI data allow us to constrain 
the beaming factor and to cross check this result with that obtained from the 
\g-ray transparency. X-ray data can shed light on the balance between the SSC 
and the ERC component, and allow us to estimate the spectral shape of the low energy 
branch of the electron energy distribution. UV data provide information about 
the Big Blue Bump (BBB) radiation, and combined with optical data tells us about the high 
energy branch of the electron distribution. Moreover, UV/optical data constrain 
the peak flux and energy of the low energy bump, determining the ratio between 
the output of the synchrotron distribution to that of the IC one.\\ 

In the following of this subsections we report the reduction of the data collected 
at different wavelengths as a result of pre-planned campaigns (GASP optical-to-radio 
and VLBI radio data) or as ToO triggered by the LAT flaring activity 
(\swf~ data). In the next Section (Sec. \ref{sec:MWRESULTS}) we discuss the 
multi-wavelength results and their connection to the LAT data. 


\subsection{SWIFT-\textit{BAT} and \textit{XRT} data} 
\label{sec:MW-BAT-XRT} 
We analyzed XRT \citep{Burrows2005,Swift} data using the \texttt{xrtpipeline} tool 
provided by the \texttt{HEADAS v6.7} software package, for data observed in photon 
counting mode. Events in the 0.3--10 keV energy band were 
extracted, selecting grades in the range 0--12, and default screening 
parameters to produce level 2 cleaned event files were applied. 
Due to the low count rate ($<2$ cts/s) we did not find any signature of pile-up effect.

We used data from the Burst Alert Telescope (BAT) on 
board the \textit{Swift} mission to derive the spectrum of \PKS~in the 14--195\,keV band. 
The spectrum is constructed by averaging 
the spectra of the source extracted over short exposures (e.g. 300\,s) and it is 
representative of the sources emission over the 
5-year time range spanned by the observations. These spectra are 
accurate to the mCrab level and the reader is referred to \cite{ajello08,ajello09a,ajello09b} for 
more details. \PKS~is bright in BAT and the approximate significance of the 
BAT spectrum used for this analysis is $\simeq$13\,$\sigma$. 

\subsection{\swu~ data} \label{sec:MW-UVOT} 
We followed the steps outlined in the UVOT User's Guide, 
to perform UVOT data reduction and analyses in all the 6 available 
filters (V, B, U, UVW1, UVM2, UVW2). 
We started from the 
raw data stored in the \texttt{HEASARC} archive and we made sure that the sky 
coordinates were updated, the modulo-8 correction was applied, duplicated FITS 
extensions have been removed, and the aspect correction was calculated. Based 
on the AGN intensity, the optimal source extraction region is a 5$''$ circle. 
The background region is an annulus with inner-outer radii of 15$''$--27$''$, 
27$''$--35$''$, depending on the filter used. In order to improve the astrometry, the NED 
position has been adjusted using the \texttt{uvotcentroid} task, and 
field of view sources have been excluded from the background region. The 
standard output of the \texttt{uvotsource} task has been used to extract the 
photometric light curves. We corrected the magnitudes for Galactic extinction 
assuming $E(B-V)_{\rm Gal}=0.097$ mag. This value was calculated from 
\cite{Schlegel1998} tables using tools provided by the 
NASA/IPAC archive\footnote{The NASA/IPAC Extragalactic Database (NED) is 
operated by the Jet Propulsion Laboratory, California Institute of Technology, 
under contract with the National Aeronautics and Space Administration.}. The 
absorption for the other filters was calculated according to the extinction 
laws of \cite{Cardelli1989}. The de-reddened magnitudes were converted into 
fluxes in physical units taking into account the zeropoints by 
\cite{PooleT.S.2008}.

\subsection{Optical Near-IR and radio observations by the GASP} 
\label{sec:MW-GASP} 
The GLAST-AGILE Support Program (GASP) is performing a long-term monitoring of 
28 $\gamma$-ray loud blazars in the optical, near-infrared, mm, and radio 
bands \citep{Villata2008,Villata2009}. The GASP has been following \PKS~
since 2007 January, and contributed to multi-wavelength studies involving 
$\gamma$-ray data from AGILE \citep{Pucella2008a,D'Ammando2009}. The optical 
and near-infrared GASP data for the present paper were acquired at the 
following observatories: Abastumani, Armenzano, Calar Alto, Campo 
Imperatore, Castelgrande, Crimean, Kitt Peak (MDM), L'Ampolla, Lowell 
(Perkins), Lulin, Roque de los Muchachos (KVA and Liverpool), Sabadell, San 
Pedro Martir, St.\ Petersburg, Talmassons, and Valle d'Aosta. Magnitude 
calibration was performed with respect to a common choice of reference stars in 
the  field of the source from the photometric sequence by 
\citet{Raiteri1998}. Conversion of magnitudes into de-reddened flux densities 
was obtained by adopting the Galactic absorption value $A_B=0.416$ from 
\citet{Schlegel1998}, consistent with the $E(B-V)$
color excess, the extinction laws by \citet{Cardelli1989}, and the 
mag-flux calibrations by \citet{Bessell1998}.

The GASP mm--radio data were taken at Medicina (5, 8, and 22 GHz), Mets\"ahovi 
(37 GHz), Noto (43 GHz), SMA (230 GHz), and UMRAO (4.8, 8.0, and 14.5 GHz). 

\subsection{VLBI data} 
\label{sec:VLBI} 
The 2\,cm VLBA / MOJAVE program (\cite{Lister2009} and references
therein) has been monitoring \PKS~at 15~GHz with the Very Long
Baseline Array (VLBA) since 1995. Method of observations, data processing and
imaging is discussed by \cite{Lister2009}. Typical resolution of these
images is about or better than 3~pc.

In addition to the 15~GHz MOJAVE VLBA monitoring, single-epoch
simultaneous multi-frequency 5-43~GHz VLBA measurements were done on
2009 April 09, in support of the first year \textit{Fermi} observations
\citep{Soko2010}.
Accuracy of flux density measurements is  dominated by
calibration uncertainties: about or less than $5\%$ at 5, 8, and 15~GHz,
about our less than $10\%$ at 24 and 43 ~GHz. 


\section{Multifrequency Results and connection with the LAT data } 
\label{sec:MWRESULTS} \subsection{X-ray and Hard X-ray data} 
\label{sec:X-ray-results} We performed the spectral analysis of \swx~data 
after grouping the photons 
to have a minimum number of 10 photons per bin, and we fitted the spectra by 
means of a \textit{photon} power-law distribution $F(E)=K~E^{-\alpha_X}$, plus a Galactic absorption 
with an equivalent column density $N_H=7.88\times 10^{20}$ cm$^{-2}$ \citep{Lockman1995}.
We extracted the X-ray spectrum for each pointing. The corresponding spectral
analysis results are reported in Tab. \ref{tab:xrt_fit}.

\begin{figure}[H]
\begin{center}
\epsfig{file=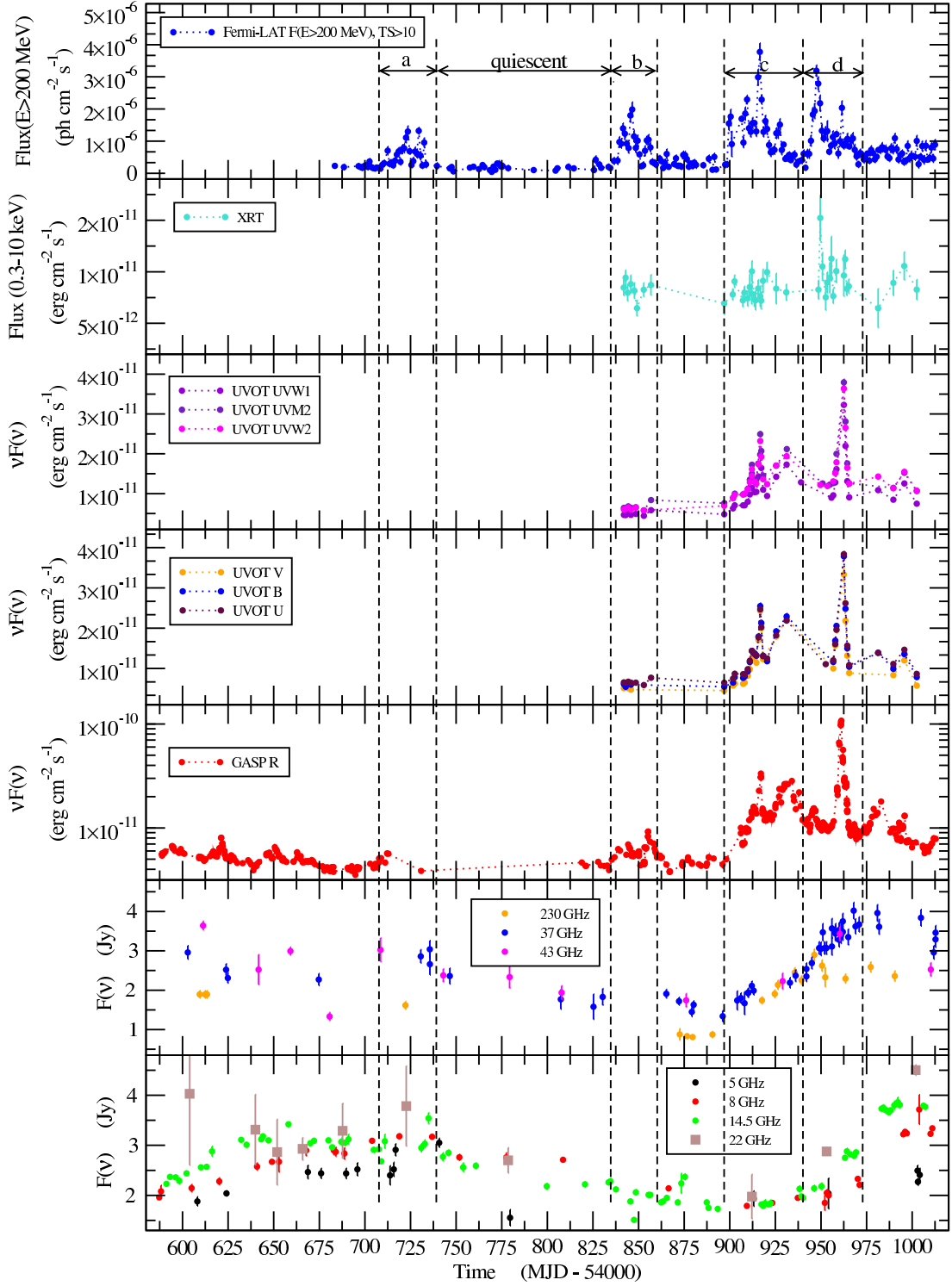,width=14.5cm,angle=-0}
\end{center}
\caption{The MW light curves, from 2008 April  to 2009 June.
The vertical dashed lines show the the four flaring episodes and
the quiescent state.}
\label{fig:MWLC}
\end{figure}

The X-ray light curve obtained from the fluxes reported in Tab. \ref{tab:xrt_fit}
shows a modest variability 
 if compared to optical and $\gamma$-ray flares (see Fig. \ref{fig:MWLC}). 
 {We prefer to present light curves in terms of $\nu F (\nu) $ to make 
easier the comparison between the various bands and the SED changes.} 
The average flux  	integrated in the $0.3-10.0$ keV range is around $10^{-11}$ erg cm$^{-2}$ 
s$^{-1}$, the lowest flux, recorded on 2009-01-16, was (6.5$\pm$0.8) $\times 10^{-12}$ 
erg cm$^{-2}$ s$^{-1}$. The highest flux, recorded on 2009-04-28, was 
(15.0$\pm$1.5)$\times 10^{-12}$ erg cm$^{-2}$ s$^{-1}$. In this case the flux increased by 
a factor of two within a day, and the spectrum reached the hardest state 
($\alpha_X=1.13\pm0.13$).  
This is the most relevant X-ray flaring episode in our data set
and looking at the MW light curve in Fig. \ref{fig:MWLC} it seems to have
no counterparts in other wavelengths. 
Since the statistics are low, we performed the spectral analysis using 
the Cash method (C-stat) \citep{Cash1979} based on the use
of a likelihood function. This method returns flux and photon index
values that are compatible with those coming from the $\chi^2$ method. 
Even if the two methods results are compatible, the significance
of this flare is low ($\simeq 2 \sigma$), so we do not investigate
possible physical implications.

During our observations the source spectrum was always hard, with a photon index ranging 
between about 1.3 and 1.6. 
The plot of the flux in $0.3-10.0$ keV range vs. the photon index (see Fig. 
\ref{fig:XRT-FULX_VS_INDEX}) is compatible with a harder when brighter trend.
Using the Monte Carlo method described in Sect. \ref{sec:Fermi-Spec-Ev},
we get a correlation coefficient $r= -0.31$ with a $95\%$ confidence
limit of $-0.55\leq r \leq -0.05 $.
This spectral trend is consistent with the same analysis performed by \cite{Kataoka2008}. 
We note also that \cite{Kataoka2008} found a soft X-ray excess in the \suzaku~ 
data, but the statistics of the individual pointings in our data set are not
sufficient to detect such a feature.

\begin{figure}[H]
\begin{center}
\epsfig{file=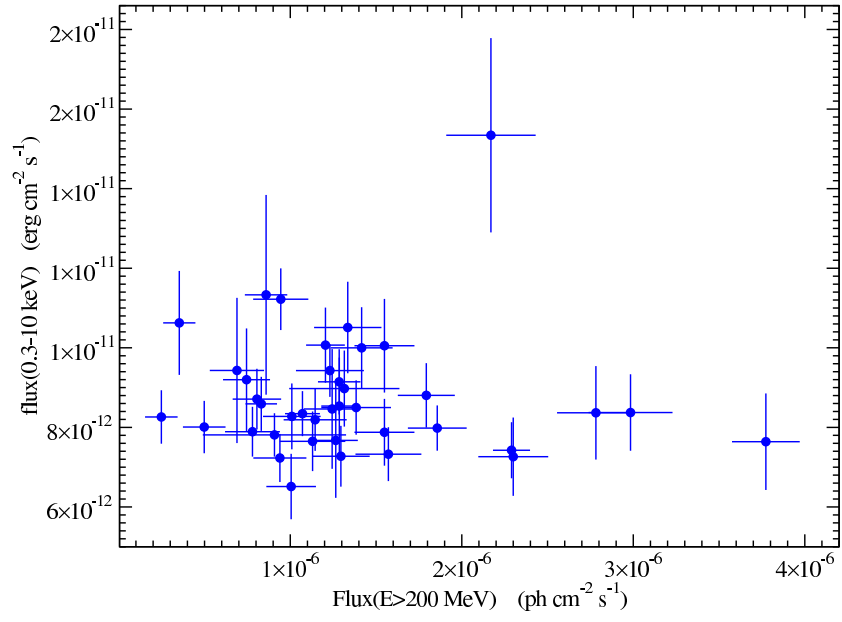,width=11cm,angle=0}
\end{center}
\caption{Scatter plot of the XRT flux in the 0.3-10 keV band vs. the \textit{Fermi}-LAT 
flux (E$>200$ MeV).} 
\label{fig:LAT-XRT-CORR} 
\end{figure}

In order to increase the statistics and to 
look for differences between the different \g-ray flares,	we produced X-ray 
SEDs averaged during the \textit{b}, \textit{c}, and \textit{d} $\gamma$-ray 
flaring intervals, and during the post-\textit{d} flaring period, reported
in Fig. \ref{fig:XRT-SEDs_AVERAGED}. These SEDs show that the average state of the 
X-ray emission was almost steady, without drastic differences between the 
flares and the post-flare integration period. We also note that a 
possible soft X-ray excess is visible in the post-\textit{b} flare averaged SED. 
The scatter plot in Fig. \ref{fig:LAT-XRT-CORR} shows no correlation between 
the XRT flux and the LAT flux. This absence of correlation is relevant to the 
understanding of the emission scenario that we will discuss in the Sect. 
\ref{sec:Discussion}. 

The 5-year integrated BAT SED is plotted in Fig. \ref{fig:XRT-SEDs_AVERAGED}.
The photon index, in the 14-150 keV band, is $1.37^{+0.08}_{-0.19}$.
Despite the long integration time of the BAT data, the photon index value
is almost compatible with the range of values observed in the XRT data in our
data set, and in other historical observations. 
This suggests that the X-ray and hard X-ray flux
and spectral shape of this source is quite stable, or at least 
that our X-ray sampling is representative of the X-ray and hard
X-ray shape on timescales of years.

\subsection{Optical/near IR and UV data results}
\label{sec:Optical and UV results}
Simultaneous \swu~ and GASP observations provide a valuable data set to study 
the low energy bump of the SED. In Fig. \ref{fig:UVOT_GASP_SED} we plot the 
SEDs obtained from UVOT and GASP data, simultaneous within a daily timescale. 
{  These data show a minimum around the frequency of 
$5\times10^{14}$ Hz, which does not seem to vary when the source is flaring. } 
The frequency of the high energy peak can be well estimated
and is close to $10^{15}$ Hz, while that of the peak at low frequency cannot be well 
established and should be roughly estimated around $10^{13}$ Hz. The UV peak, 
as in many quasars, is likely due to the the BBB that 
usually is understood as thermal emission from the accretion disk surrounding 
the black hole. Assuming that the disk luminosity is steady or 
slowly variable compared to the synchrotron emission, we expect that the UV excess 
gets less and less evident, as the optical flux increases. To test this 
scenario, in Fig. \ref{fig:UVOT_GASP_SPEC_EVOL} we plot the ratio of the 
flux in the highest energy UVOT filter (UV\textit{W2}) to the flux in the 
\textit{B} UVOT filter, as a function of the optical $R$ flux. 
This plot shows that the UV spectrum gets harder when the optical \textit{R} flux 
is lower. This trend is consistent 
with the UV excess decreasing as the \textit{R} flux is increasing, as expected in 
a scenario in which the UV bump originates in the thermal emission of the 
accretion disk. 
Despite this trend, we note that even in the highest state plotted in Fig. 
\ref{fig:UVOT_GASP_SED}, the corresponding UV bump is still prominent. 
For other FSRQs, such as 3C~454.3, the BBB is not visible 
during the high optical states 
\citep{Giommi2006,Villata2006a}, since it is dominated by the synchrotron flux, as
we would expect here in the case of \PKS. In the panels \textit{a} and \textit{b} of 
Fig. \ref{fig:UVOT_LC} UV light curves for all the 6 filters are shown, and in 
the panel \textit{c} we plot the ratio of the UVOT \textit{V} filter flux to 
the UVOT \textit{W2} filter. The trend of the hardness ratio is clearly 
anticorrelated with that of the fluxes. Again, this supports the BBB scenario, 
with the UV spectrum harder when the BBB is more evident, namely when the 
synchrotron flux is lower. 

In Fig. \ref{fig:LAT-UVOT-CORR} we report the scatter plot of the \textit{Fermi}-LAT flux 
F(E$>200$ MeV) vs. the UVOT $\nu$F($\nu$) in the UV\textit{W2} filter, to check for a 
possible significant correlation between the \g-ray and UV fluxes. The 
correlation coefficient of the logarithms of the UV and $\gamma$-ray fluxes,
obtained through the Monte Carlo method described in Sect. \ref{sec:Fermi-Spec-Ev}, 
is $r=0.2$ 
with a $95\%$ confidence interval of $0.05 \leq r \leq 0.34$. If we exclude
the point with the lowest \g-ray flux, the correlation coefficient is 
$r= 0.05$, suggesting that the overall correlation is not
significant.

This lack of correlation hints that the spectral evolution of the UV spectrum 
results from the contamination of the high energy branch of the synchrotron emission, 
and that a change in the BBB luminosity is not the driver of the \g-ray luminosity 
variations. We will discuss this in Sect. \ref{sec:Discussion}.

The optical (\textit{R} band) and near-IR (\textit{J, H}, and \textit{K} bands) 
observations span the period from 2008 May until the end of 2009 June. In Fig. 
\ref{fig:GASP_OPT_LC} we report the light curves  of the optical
data. The best sampled band is the \textit{R} one, with more than 600 
observations. The optical flaring activity increased dramatically after mid 
March 2009, corresponding to the $\gamma$-ray flare \textit{c}. On 2009-05-08 
the source was very  bright with an $R$  magnitude of $13.60\pm0.02$ \citep{LarionovV.M.2009a}, 
and  in two days it reached its historical peak at $R=13.07 \pm 0.02 $, 
about $1.3$ magnitude brighter than the previous record level, observed on March 27 
\citep{LarionovV.M.2009} of the same year. 
{
This dense monitoring is fundamental to understand the correlation between 
the optical and the \g-ray activity, both for the long-term trends and the single flares. 
In Fig. \ref{fig:LAT-OptR-CORR} we show the scatter plot of the \textit{Fermi}-LAT flux  
F(E$>200$ MeV) vs. the optical $\nu$F($\nu$) in the $R$ filter.
The correlation coefficient of the logarithm of the 
optical and $\gamma$-ray fluxes, evaluated through the Monte Carlo
method described in Sect. \ref{sec:Fermi-Spec-Ev}, is $r= 0.42$ 
with a $95\%$ confidence interval $0.36\leq r\leq 0.46$,  higher 
than that found for the UV band. This finding hints
that the driver of the flaring activity can be a change in the high energy
branch of the electron distribution. We will investigate this scenario 
more accurately  in Sect. \ref{sec:Discussion}.
}{
Despite the statistical significance of this result, there
is an evident dispersion in the scatter plot, that could be related
to the inter-band time lags. 
Indeed, temporal lags could be related to the internal source photon absorption, 
to the cooling time of the radiating particles, or to inhomogeneities 
in the emitting region. 
To search for  possible time lag 
between the optical (\textit{R} band) and the $\gamma$-ray band, we used the discrete 
cross correlation function (DCCF) method. Taking into account the whole 
data set, the DCCF analysis returns a lag of $13\pm1$ days, with
the $\gamma$-ray band leading the optical one (see Fig. \ref{fig:DCCF1}, top panels). 
Since the \textit{R} flux increased dramatically after the end of 2009 February, we 
divide the data set in two Sections, before and after MJD 54890 (corresponding to 
2009-02-28). The DCCF results, plotted in the middle panels of Fig. 
\ref{fig:DCCF1}, show that the 13 days delay is still apparent. 
The same result holds if we analyse individually the flaring sequences \textit{d} 
and\ \textit{b} (Fig. \ref{fig:DCCF1}, bottom panels).  
Looking at the flare light curves it is clear that the 13 day lag is caused 
by a change in the relative flux of the two bands such that the $\gamma$-ray 
flux is stronger in the first part of the flare and the R-band flux gets 
stronger in the second half. We plot in the lower panel of Fig. 
\ref{fig:LAT-OptR-SHIFTED} the LAT light curve above 200 MeV, and in the upper panel the 
optical light curve, backward shifted by 13 days: 
the four $\gamma$-ray flares \textit{a, b, c, d}, seem to have
an optical counterpart. 
In particular, the peaks of the bright events in May would  be very close in time to those 
in the LAT daily light curve at MJD $\simeq$ '948 (see Fig. \ref{fig:LAT_SING_FLARES}). 
To check the effect of this lag,  we evaluate the correlation between the logarithms 
of the optical and  $\gamma$-ray flux  after applying the above time shift
(see Fig. \ref{fig:LAT-OptR-CORR-SHIFTED})  and find   $r = 0.62$ with a 95$\%$
confidence interval $0.57 \leq r \leq 0.66 $. 
{The values of the correlation coefficient and of its 95$\%$
confidence interval, after applying the 13 days time shift, 
are significantly higher than those obtained without the time shift. 
Moreover, Fig. \ref{fig:LAT-OptR-CORR-SHIFTED}  and Fig. \ref{fig:LAT-OptR-CORR} 
show that time shift reduces the dispersion in the scatter plot.}}

{ This 13 day correlation between the optical and $\gamma$-ray emission 
appears  in the different outbursts during
this flaring activity but we do not have any indication that it is a characteristic 
behavior related to the temporal and energetic evolution
of the flares. Only through a long-term simultaneous \g-ray and 
optical observations  may we understand better the actual level
of randomness of this correlation and its possible physical meaning.}

\subsection{Radio results}
\label{sec:mm-Radio results} 
Radio data have a much lower sampling w.r.t. optical and $\gamma$-ray, but at 14.5 
GHz and 43 GHz it is possible to follow the overall trend of the radio flaring 
activity. Even if radio fluxes do not show a correlation with the optical and 
 the \textit{Fermi}-LAT flaring trend, it is possible to identify a large radio flare, 
starting at about MJD 54900 (see Fig. \ref{fig:MWLC}). 
This radio flare, visible  at 14.5, 43, and 230 
GHz, seems to starts quasi-simultaneously with the $\gamma$-ray flare \textit{c} 
at 43 GHz, and keeps increasing until the end of flare \textit{d}. The 230 GHz 
light curve shows a structure similar to a plateau, starting when the $\gamma$-ray 
flare \textit{ c} has ended, and a possible plateau is present also at 43 GHz, 
starting at the end of flare \textit{d}. The 14.5 GHz fluxes seem to lag behind 
the 43 and 230 GHz. Analysis in the radio band is much more complex than optical 
and $\gamma$-ray data, because of the synchrotron self-absorption and the longer cooling 
times, and the possibility of the overlapping of different flares.
 
\begin{figure}[H]
\begin{center}
\epsfig{file=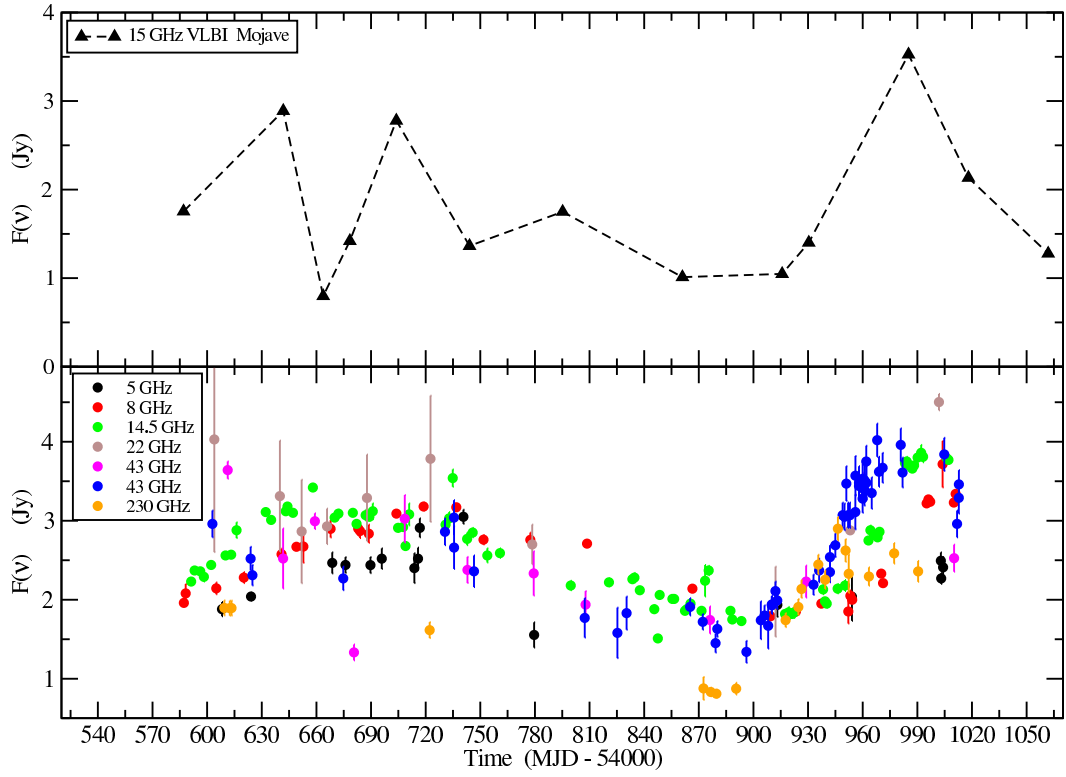,width=14cm,angle=0}
\end{center}
\caption{{\it Upper panel:} 
The 15 GHz MOJAVE VLBA parsec-scale core light curve. 
{\it Lower panel:} GASP radio light curve. } 
\label{fig:LC-GASP-VLBI} \end{figure}

In  2009 June-July the integrated parsec-scale flux density had reached
its historical maximum since 1995 ($\simeq$ 4 Jy), as the 15~GHz MOJAVE VLBA
measurements show. The VLBI core flux density has
also shown the highest value (see Fig. \ref{fig:LC-GASP-VLBI}, upper panel). 
If this major radio flare, observed in the core of the parsec-scale jet, 
is connected to the huge $\gamma$-ray flares which have happened in the 
first half of 2009, this determines the source of the high energy
emission to be located around the base of the parsec-scale jet following 
causality arguments, as suggested by \cite{Kovalev2009}. The delay between 
the peaks of light curves in the $\gamma$-ray and radio bands can be, at least 
partly, explained by the synchrotron self-absorption of the emission at 
radio frequencies.

\begin{figure}[H]
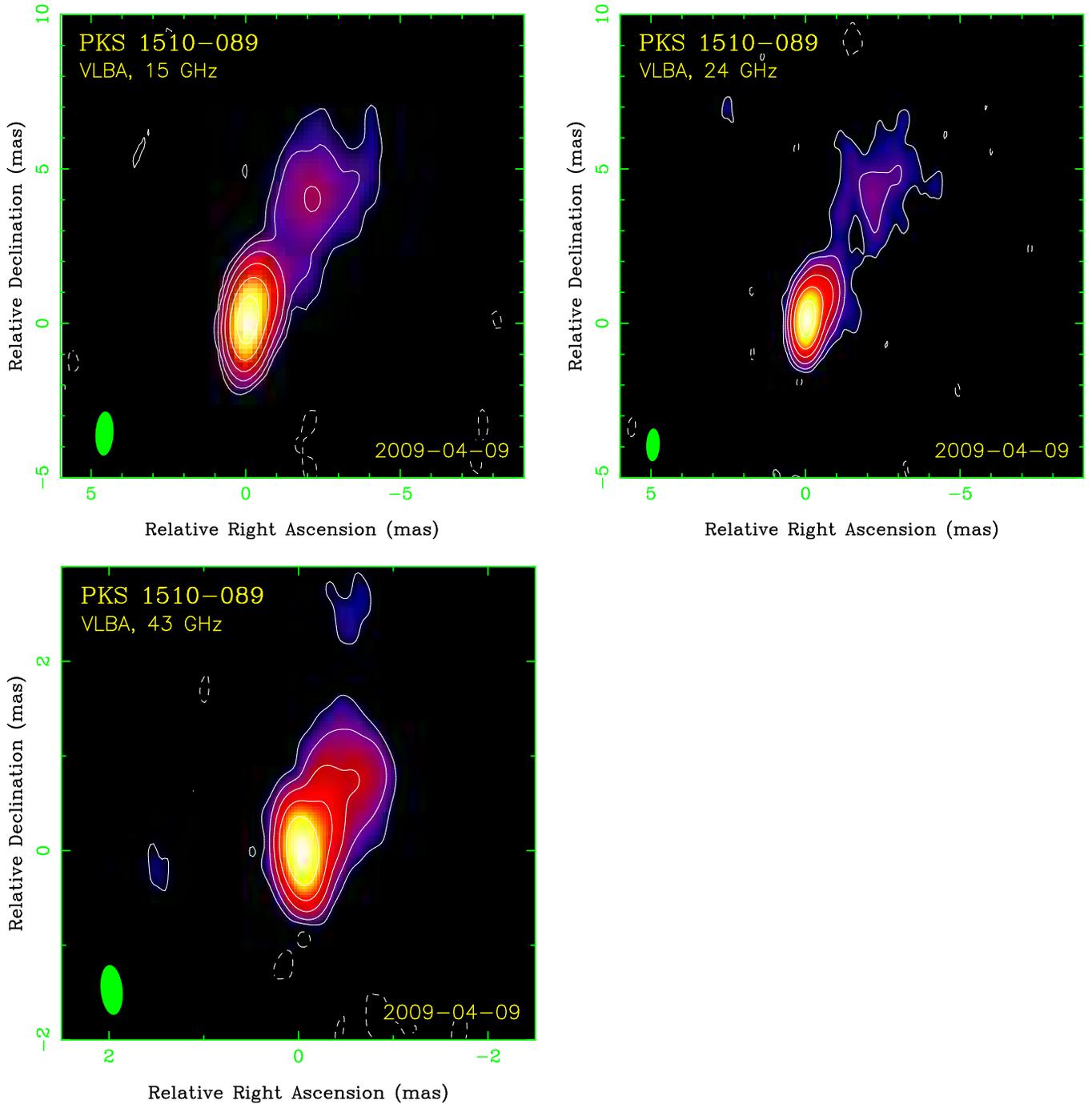

\begin{center}
\begin{tabular}{ll}
\epsfig{file=fig21a.eps ,width=9.cm,angle=-90}&
\epsfig{file=fig21b.eps ,width=9.cm,angle=-90}\\
\epsfig{file=fig21c.eps ,width=9.cm,angle=-90} 
\end{tabular}
\end{center}
\caption{Stokes I CLEAN images of PKS~1510-089 observed by VLBA on 2009 April 9,
at 15, 24, and 43~GHz. The lowest contour and peak intensity are
0.7~mJy/beam and 1.47~Jy/beam (15~GHz), 0.7~mJy/beam and 1.56~Jy/beam
(24~GHz), 2~mJy/beam and 1.85~Jy/beam (43~GHz). Contours are plotted
with a step $\times 4$. Natural weighting of visibility data is used,
HPBW beam size is shown in the lower left corner. Angular size of 1~mas
corresponds to 5~pc.} 
\label{fig:VLBI_IMAGE} 
\end{figure}

\begin{figure}[H]
\begin{center}
\includegraphics[width=14cm,angle=0]{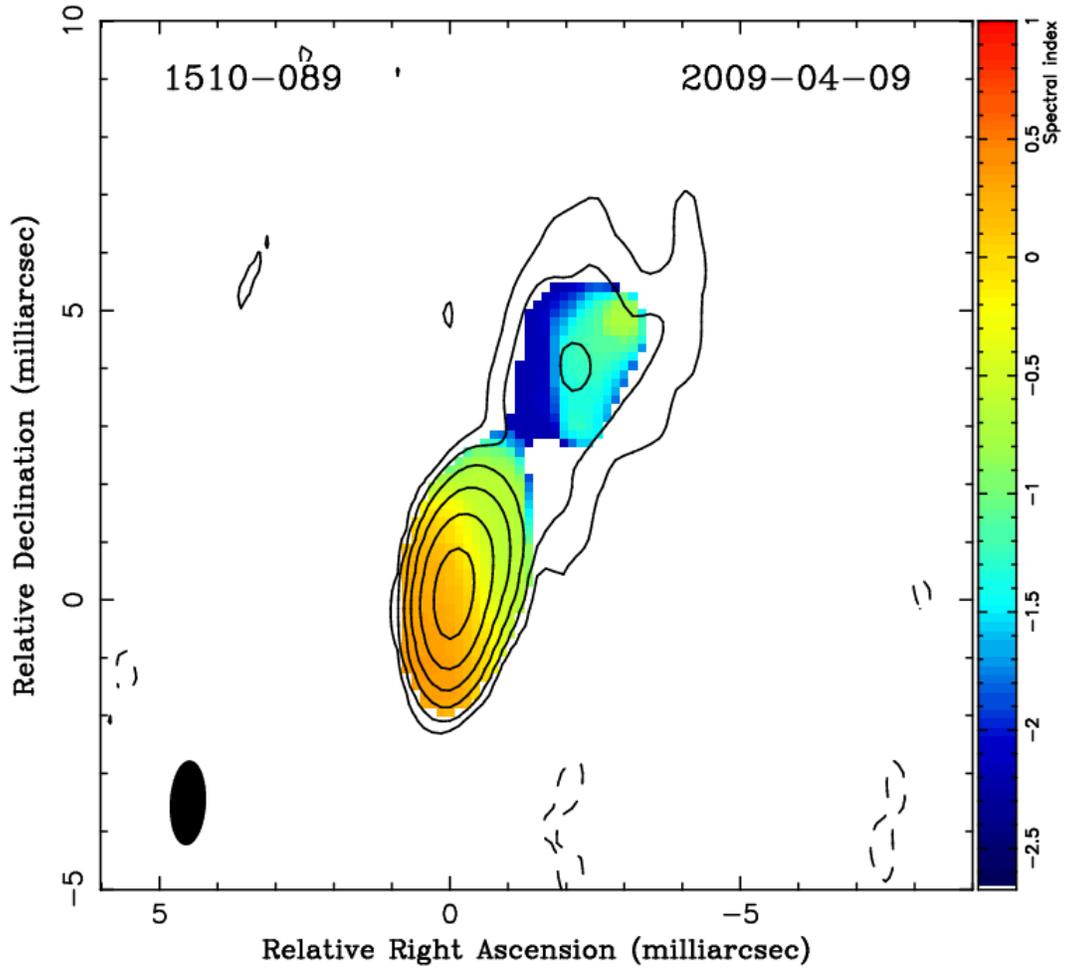}
\end{center}
\caption{Spectral index $s_r$ ($F(\nu)\propto\nu^{s_r}$) map between 15.4 and
23.8~GHz (shown in color) of \PKS~as observed by the VLBA on 2009 April 9. 
The overlaid contours represent total intensity at
15.4~GHz (see Figure~\ref{fig:VLBI_IMAGE} for details). The spectral index map
was smoothed by a median filter with a 0.6~mas radius. }
\label{fig:VLBI_Spectrum} 
\end{figure}

In Fig. \ref{fig:VLBI_IMAGE} we present the
three highest frequencies Stokes~I parsec-scale images,
from the 5-43~GHz VLBA measurements performed on  2009 April 9. The size of
the bright parsec-scale core at 24 and 43 GHz is estimated to be about
60-70~$\mu$as or 0.3-0.4~pc. The core shows a flat radio spectrum
(see Fig.~\ref{fig:VLBI_Spectrum}) indicative of a synchrotron self-absorbed
region, while the first well resolved jet feature is already optically
thin radio spectral index $s_r=-0.9$ ($F(\nu)\propto\nu^{s_r}$).

The highest apparent speed of a component motion in the jet of
\PKS~observed at 15~GHz by \cite{Lister2009a} is
$v_\mathrm{app}=24 c$ which makes its jet highly relativistic and
Doppler boosted, a typical case for $\gamma$-bright blazars
\citep{Lister2009b}. 
\cite{Marscher2010} report on two new knots, observed in VLBA
images at 43 GHz. The first with an apparent speed of $24\pm2c$,
passed the core on MJD $54674.5 \pm 20$ (2008-07-27). The second
knot passed the core on MJD $54958.5\pm 4$ (2009-05-07) 
which is consistent, within the estimated uncertainties, 
with the huge optical flare observed on 2009-05-08. 
A feature, coincident to the second knot, is also seen to emerge in the MOJAVE 15 GHz VLBA 
images from 2009 June  through  2009 December. The fitted speed at 15 GHz 
($675 \mu \mathrm{as \; y^{-1}}$) is somewhat slower than the \cite{Marscher2010} speed 
($970 \pm 60 \; \mu \mathrm{as\; y^{-1}}$), {but this
could be due to blending with a third feature seen in the 43 GHz data
to emerge sometime in June 2009 \citep{Marscher2010b}. Despite the
high resolution of the 43~GHz images, a more accurate ejection date
for the latter feature could not be obtained due its rapid angular
evolution and close proximity (0.1 mas) to the feature ejected in 2009 May.}

\section{Interpretation and discussion}
\label{sec:Discussion} 
\subsection{Physical interpretation of $\gamma$-ray and MW data }
\label{sec:gamma-ray-phys}
\PKS~was one of the brightest and most active blazars observed by \latf~ 
during the first year of survey. 
Flaring episodes with timescales from weeks to months in addition to rapid 
and intense outbursts, were observed both at optical and \g-ray energies. 
The estimated isotropic luminosity above 100 MeV during the 
flare \textit{c} was about 8$\times 10^{47}$\lum. 
This value represents the brightest flare-averaged state of the source. 
The brightest daily time-resolved luminosity, recorded during the flare 
\textit{c} on 2009-03-26, was of $\simeq 2\times 10^{48}$\lum. 
During the flare \textit{d}, variability timescales, down to a fraction of a day, 
were observed both at optical and LAT energies.
 
The rapid 
variability and the powerful \g-ray luminosity raise the problem of the pair 
production opacity. 
Indeed, without beaming effects, the source size estimated from the observed variability 
timescale ($R_{rad}=c\Delta t$/(1+z)), make the source opaque to the 
photon-photon pair production process, provided that \g-ray and X-ray photons 
are produced cospatially.

If the component emerged on 2009-05-07, observed also in the MOJAVE 15 GHz VLBA,
is related to the flares \textit{a} and \textit{b}, then we can use the value of 
the beaming factor derived from the motion of radio knots, namely
$\delta\simeq\Gamma\simeq\beta_{app}\simeq 21$.
This estimate is compatible with other VLBI estimates \citep{Homan2002},
and is slightly larger than an alternative estimate based on the variability 
observed at 22 and 37 GHz, $\delta_{var}= 16.7$ \citep{Hovatta2009}. 
Taking into account the most rapid timescale estimated in Sect. \ref{sec:Fermi-time}, 
$\Delta t\simeq 0.25$ days, the actual emitting region size results of the order of 
$R_{rad}\leq \delta \Delta t c /(1+z)\simeq 1\times 10^{16}$ cm. 
Using the above fastest timescale 
and the quasi-simultaneous observed X-ray flux $F_X\simeq 8\times 10^{-12}$\flux 
(observed at a typical frequency of $10^{18}$ Hz), one can impose a limit on the 
minimum value of the beaming factor resulting in a source transparent to the 
photon-photon annihilation process \citep{Maraschi1992,Mattox1993,Madejski1996}. 
Combining the source size from Eq.~\ref{eq:size_constr}, the X-ray photon energy 
in the source frame, and the intrinsic X-ray luminosity 
with the optical depth expression, we get 
a model independent estimate of $\delta \gtrsim 8$. 
This lower estimate indicates that the values derived from the VLBI images are 
well above the pair production transparency limit.

X-ray fluxes, spectral indices and trends in our data set are compatible with 
those reported in previous analysis with several X-ray telescopes  \citep[\sax,~ \asca,~ 
\suzaku;][]{Kataoka2008}. 
The harder when brighter trend and the lack of X-ray/\g-ray correlation are 
useful to constrain both the low energy tail of the electron distribution and the 
emission scenario. 
The typical value of the soft-to-hard-X-ray photon index $\alpha_X$ is close to 1.4,
with a quite narrow dispersion ($\simeq 0.1$), and is similar to that observed in 
FSRQs with $z>2$ \citep{Page2005}. 
Since in FSRQ objects the X-ray band samples the low-energy tail of the ERC component,
the X-ray energy spectral index ($s_X$) constrains the slope $p$ of the low-energy tail of 
the electron distribution in the range 1.6-2.0  \citep[$p=2 s_X-1$, e.g.][]{Ryb1979}.

\begin{figure}[H]
\begin{center}
\epsfig{file=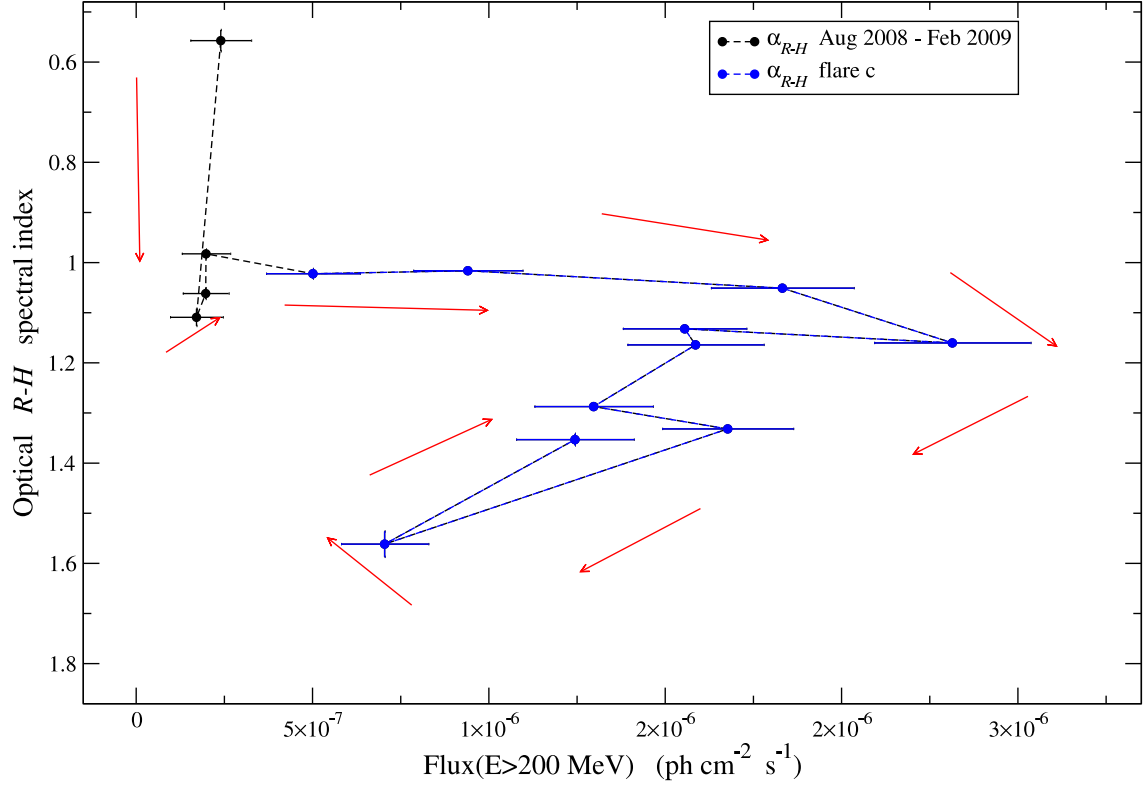,width=15cm,angle=-0}
\end{center}
\caption{Optical $R-H$ spectral index vs. the LAT Flux above 200 MeV. 
The black points refer to the 2008 August  to 2009 February period,
and are poorly sampled to be used to investigate the optical/\g-ray connection. 
The blue points overlap mainly the flare \textit{c}, in particular the subflare 
peaking at about MJD 54909 (see fig. \ref{fig:LAT_SING_FLARES}, panel d). The
red arrows shows the chronological sequence.}
\label{fig:OPT-RH-Index-vs-LAT-Flux}
\end{figure}

The high energy spectral index of the electron distribution can be
estimated from the Optical/UV spectral shape. 
In Sect. \ref{sec:Optical and UV results} we showed that the Optical/UV spectral 
index depends on the relative contamination between the synchrotron component and the BBB emission.
A reliable estimate of the synchrotron spectral index could be achieved from the optical data, 
but, unfortunately, our spectral coverage with two or more simultaneous frequencies
is only in a limited time window. 
In Fig. \ref{fig:OPT-RH-Index-vs-LAT-Flux} we plot the spectral index in the \textit{R} 
to \textit{H} bandpass vs. the \g-ray flux integrated above 200 MeV. 
If we take into account the blue points which refer to the subflare 
peaking at about MJD 54909 (see fig.\ref{fig:LAT_SING_FLARES}, 
panel d), we note that the \g-ray flux increased by about a factor 
of 2.5 with the optical spectral index almost constant around $\simeq 1$. 
This value of the optical spectral index corresponds to 
an electron energy distribution index of about 3. 
We note that the steadiness of the optical $R-H$ spectral index, during this 
\g-ray subflare, is consistent with the spectral evolution of the
\g-ray emission. Indeed, extracting the \g-ray photon indices simultaneous 
within one day with the optical ones, we note 
that also these values are almost stable around the value of 2.4. 
The decay of the flare occurred with a 
significant spectral softening, consistent with a cooling dominated regime, 
with a corresponding electron spectral index in the range between 3.5 and 4.0. 

The observed multi-wavelength SEDs of \PKS, reported in Fig. \ref{fig:MWSED_BKN}, 
show that during the flaring state, the IC peak dominated over the synchrotron one by more 
than one order of magnitude. 
Indeed, even if both the synchrotron and IC peak frequencies are not sampled in our data set, 
we can estimate their typical peak flux {by extrapolating the LP best fit model for 
the $\gamma$-ray data, and by fitting the radio, optical, and UV SED points by means 
of cubic function.} In the case of the synchrotron component, simultaneous mm and optical/UV 
data indicate that the peak flux should not be higher than a few times $10^{-11}$\flux. 
In the case of the IC component, since the spectrum is mildly curved with the peak 
energy below the threshold of our analysis (200 MeV), the peak flux must be 
a few times $10^{-10}$ \flux.
The presence of a pronounced BBB and the Compton dominance of about 10, during 
the flaring states, suggest that the contribution from photons originating outside 
the jet is appropriate to model the broadband SED.

\begin{figure}[H]
\begin{center}
\begin{tabular}{c}
\epsfig{file=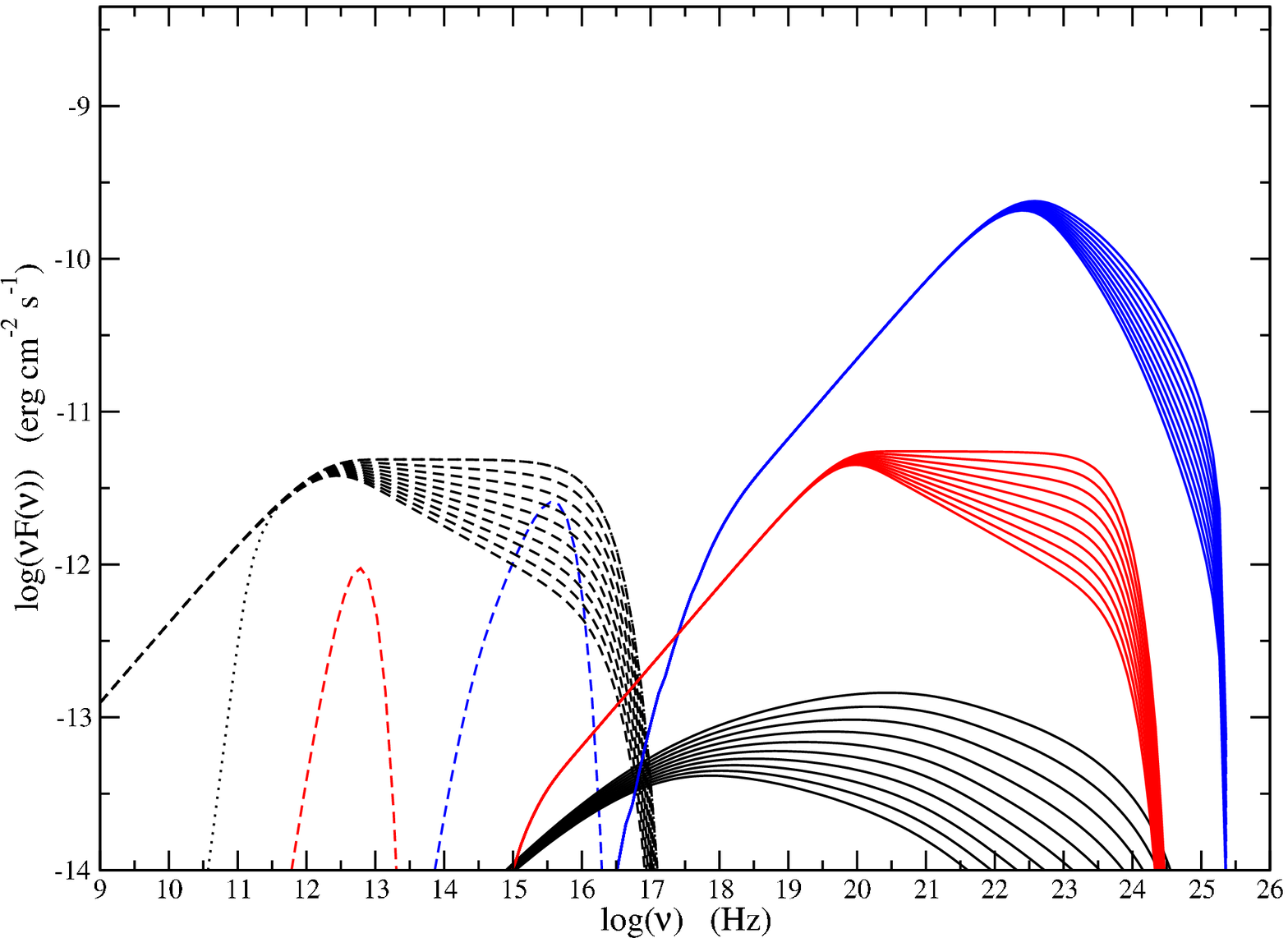,width=11.5cm,angle=0} \\
\epsfig{file=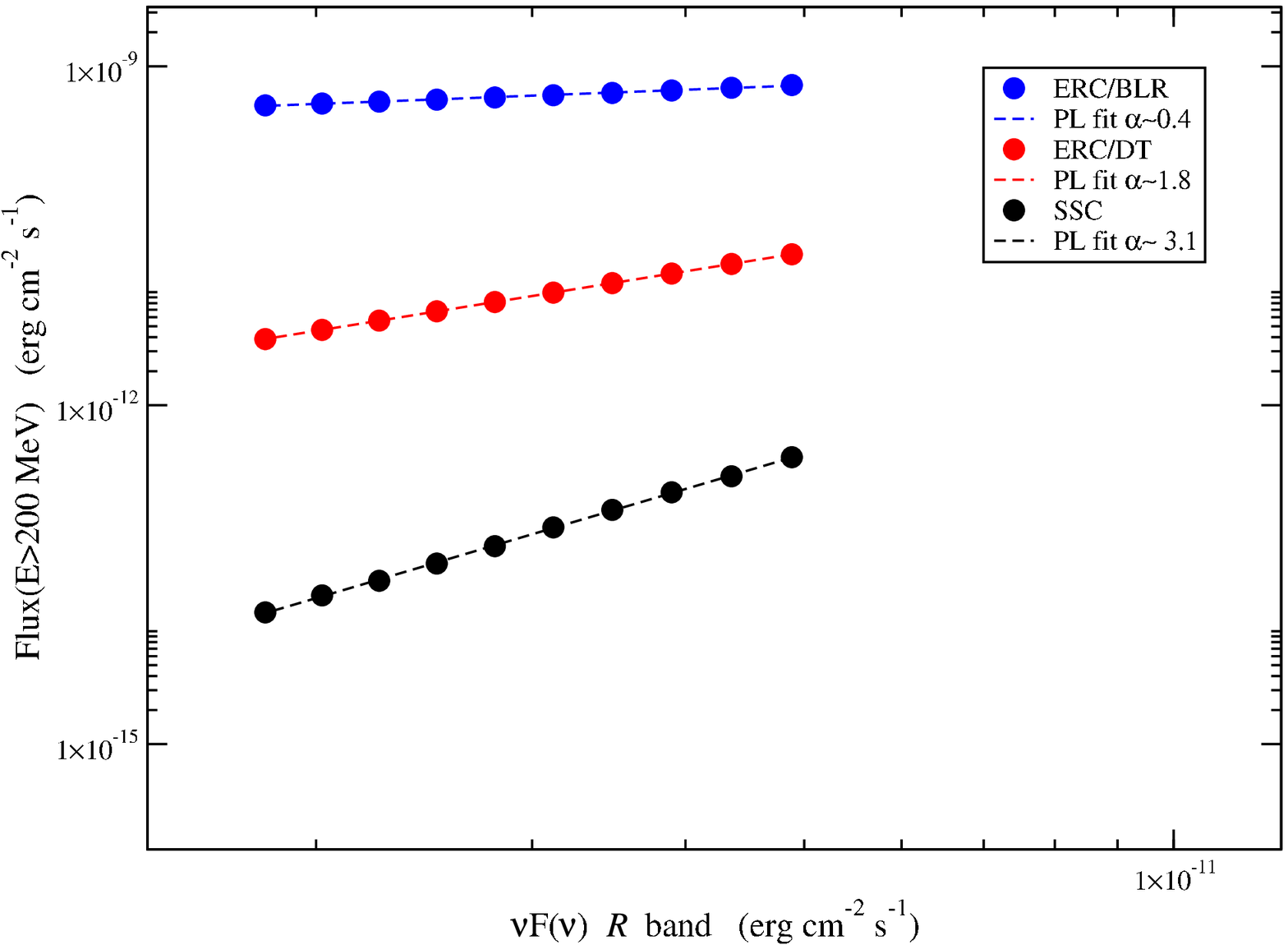, width=12cm,angle=0}\\
\end{tabular}
\end{center}
\caption{ 
{\it Upper Panel}: the SEDs for the different values of the high energy electron 
distribution index. {The black dashed lines represent the synchrotron emission 
without self absorption. The black dotted line represents the synchrotron
self adsorbed emission. The black solid lines correspond to SSC emission.
The red and blue dotted lines correspond to the dusty torus
and BBB emission respectively. The red and blue solid lines correspond to the ERC/DT
and ERC/BLR emission respectively.}
{\it Lower Panel}: optical/\g-ray flux correlations for ERC/BLR
 (blue solid circles), ERC/DT (red solid circles), and SSC (black solid circles); dashed 
 lines are the power law relations.
}
\label{fig:R-vs-LAT-simulate-Trend}
\end{figure}

The absence of X-ray/\g-ray flux correlation hints that the ERC flux variations 
depends on a change in the high energy spectral index of the electron distribution,
instead of a change in the external radiation field.  
In Sect. \ref{sec:Optical and UV results} we noted that the lack of \g-ray/UV 
correlation suggests that BBB variations are excluded as a main driver of the \g-ray 
variability. 
We tested the change of electron distribution by studying the correlation between 
the logarithms of the \g-ray and of the Optical \textit{R} fluxes (see Fig. \ref{fig:LAT-OptR-CORR}). 
The relation between the $\gamma$-ray and optical flux, fitted by means of power law, 
returns an exponent of $\sim$ 0.5 (or $\sim$ 0.6 if we use the 
13-day shifted optical light curve discussed in Sect. \ref{sec:Optical and UV results}). 
Despite the large scatter, this value is very interesting, because synchrotron and IC fluxes 
correlate very differently in the case of SSC, ERC/BLR, and ERC/DT. 
Moreover, the correlation depends on the low and high 
energy range chosen respectively for the synchrotron and IC component \citep{Katar2005}.  
To have an estimate of the power-law exponent for this correlation, we reproduced 
numerically a set of SEDs comparable to that observed for \PKS, using a well tested
code \citep{Tramacere2007PhD,Tramacere2009,Tramacere2003}. 
We calculated an ERC-dominated model, where the \g-ray emission is dominated by 
the BLR contribution and the electron distribution is a BPL. 
We fixed all the parameters and we increased the high energy index of the electron 
distribution from -3.5 to -2.5 to evaluate the correlation between the 
optical energy emission ($\nu F(\nu)$ at $10^{14}$ Hz) and the integrated 
$F(E>200$ MeV$)$ \g-ray flux, taking separately into account the three IC 
components SSC, ERC/BLR and ERC/DT, as shown in Fig. \ref{fig:R-vs-LAT-simulate-Trend}. 
The resulting flux relations were fitted by simple power laws and the resulting 
exponents were 3.1 for the SSC, 1.8 for ERC/DT and 0.4 for ERC/BLR.
The last one was the only one found to be consistent with that observed for 
\PKS~($\simeq$ 0.5-0.6) and therefore, we can disfavor both SSC and ERC/DT.

The SEDs of the ERC/BLR model in the upper panel of Fig. \ref{fig:R-vs-LAT-simulate-Trend}
have a curvature more pronounced than the ERC/DT ones. 
This is due to the Klein-Nishina cross section, because UV photons emitted from the 
accretion disk and reflected towards the jet by the BLR are blue shifted roughly by 
a factor of $\Gamma $.
For $\Gamma \simeq$ 10, the typical energy of these photons in the emitting 
region rest frame is then $\simeq$ $10^{16}$ Hz, hence the IC scattering with the 
electrons with $\gamma \simeq$ 1000 occurs under a mild KN regime. 
The resulting smooth curvature has a value of $\simeq 0.1$ that is compatible with
the observed one.

\subsection{ Spectral energy distribution modelling and jet energetics} 
\label{sec:SED-Model}
\begin{figure}[H]
\begin{center}
\epsfig{file=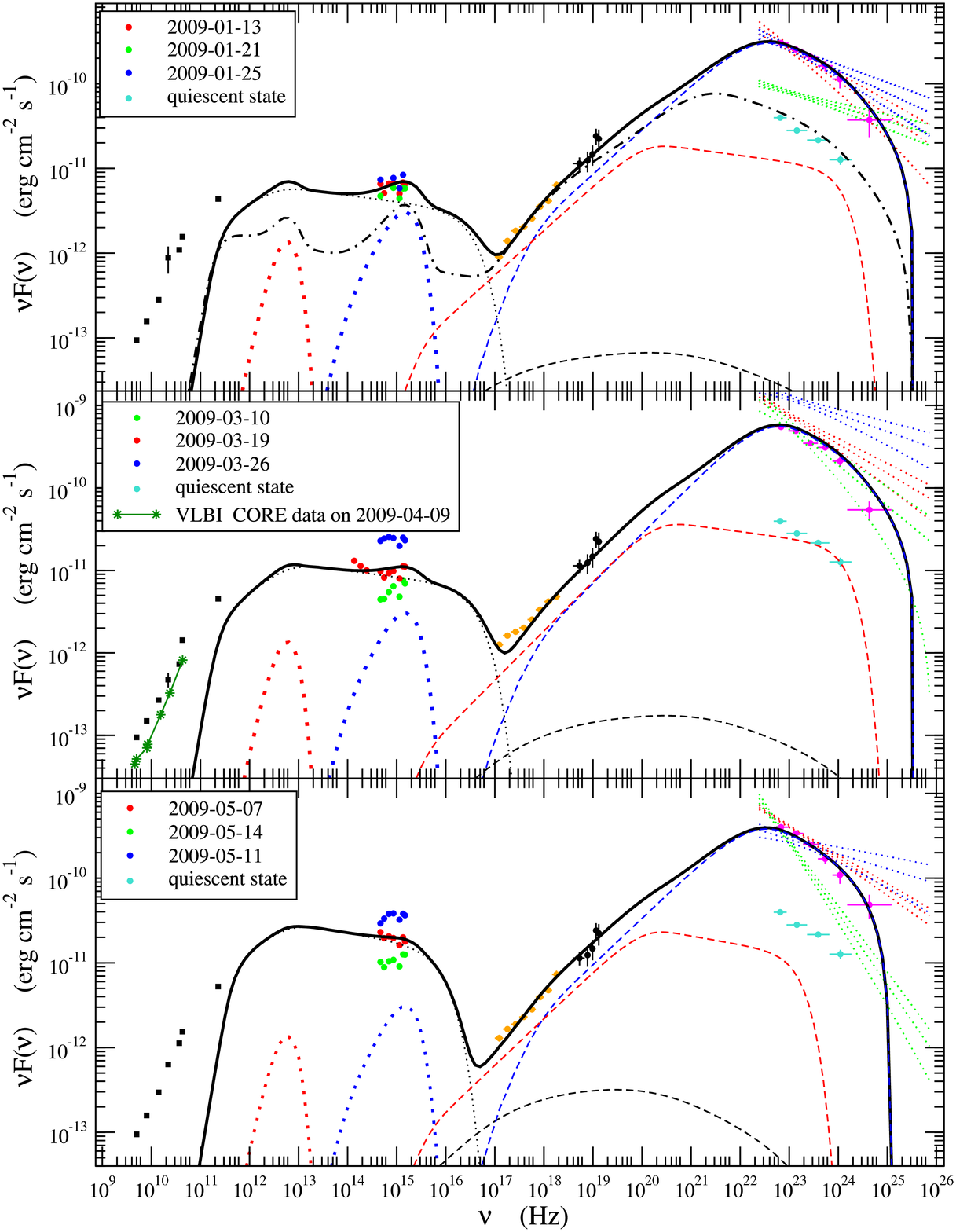,width=11cm,angle=-0}
\end{center}
\caption{Red solid circles correspond to simultaneous
Optical/UV data representing the intermediate state during the \g-ray
integration period. Blue and green circles correspond to the highest and
lowest flux state, respectively, observed during the same
integration period. The \swx~ data (orange points) are integrated during the
same interval of the LAT data, and the \swb~ data represent the 5-year
fluxes discussed in Sect. \ref{sec:X-ray-results} . 
The black squares represent GASP radio data, integrated during
the flares.
The cyan points correspond to the LAT quiescent state SED.
The magenta points represent the \g-ray SED integrated during 
the flare. The blue, green, and red dotted lines represent 
the PL best fit of the \g-ray data, for a daily integration 
and simultaneous to the Optical/UV data with the same color. 
{The thin black dotted line represents the flaring synchrotron emission. The thin black dashed line corresponds 
to flaring SSC emission. The red and blue thick dotted lines correspond to the dusty torus
and BBB emission respectively. The red and blue thin dashed lines correspond to the ERC/DT
and ERC/BLR flaring emission respectively. The solid black thick line represents the sum of all the 
flaring model components. The thick dot-dashed line represents the sum of all the model components for the
quiescent state alone.}} 
\label{fig:MWSED_BKN}
\end{figure}

We attempt a leptonic ERC/BLR-oriented SED modelling. We aim to reproduce the SEDs 
for the 3 flares with simultaneous data from radio to \g-ray energies,
namely flares \textit{b,c,} and \textit{d}. Moreover we try also to fit the 
quiescent state. 
Although variability timescales reached values of a fraction of day, LAT data 
required longer integration times to produce a SED enough good for spectral modelling.
The intermediate fluxes of the synchrotron and BBB components, observed
during the \g-ray integration period (red filled circles
in Fig. \ref{fig:MWSED_BKN}), are used in the fitting procedure as 
representative of the flare-averaged state of the low energy bump.

Since the correlation between the \g-ray and the optical
fluxes (see. Sect. \ref{sec:gamma-ray-phys} and Fig. \ref{fig:R-vs-LAT-simulate-Trend})
favours an ERC/BLR scenario, we assume that the dissipation zone is in the sub-pc scale. 
This is also consistent with the mild curvature observed in the \g-ray spectra.
Indeed, as shown in the previous Section, the ERC/BLR process 
occurs under the KN regime leading to the curved MeV/GeV spectral shape 
that matches the one observed in the \textit{Fermi} spectra.\\

We assume a jet viewing angle of $\theta=2.5^{\circ}$ for both the flaring and the quiescent states. 
During the flaring states, we choose a bulk Lorentz factor in the range [14-16], resulting in a beaming 
factor range of [20-21.5] that is compatible with the VLBI observations. During the quiescent state, we use 
a bulk Lorentz factor of 12.0, corresponding to a beaming factor of about 18.8. 

As a further step, we estimate the accretion  disk  physical characteristics using the UV data. 
We use the UV observations during the lowest synchrotron state of our dataset as an estimate for 
the upper limit of the accretion disk luminosity ($L_d$). According to the observed UV flux and to the 
luminosity distance, we set a reference value of $L_d \simeq 3\times 10^{45}$ erg/s. 
We model the accretion disk, following the prescription in \cite{King2008} and rearranging 
the expression as in \cite{Ghisellini2009},  using a multi-temperature blackbody with a temperature profile given by:

\begin{equation}
\label{eq:T_disk}
T_{disk}^4(R)=\frac{3 R_S L_d }{16 \pi \epsilon \sigma_{SB} R^3} \Big[1-\Big(\frac{3 R_S}{R}\Big)^{1/2} \Big],
\end{equation}
where $\sigma_{SB}$ is the Stefan–Boltzmann constant, $R_S=2GM_{BH}/c^2$ is the Schwarzschild radius for 
a black hole (BH) mass $M_{BH}$, $3R_S$ is 	the last stable orbit in the case of 
Schwarzschild BH, and $\epsilon$ is the accretion
efficiency that is linked to the bolometric luminosity and to the
accretion mass rate $\dot{M}$, by $L_d=\epsilon \dot{M}c^2$. 
We assume that the radiative region of the disk extends 
from $\simeq 3 R_S$ to $\simeq 500 R_S$. Since $T_{disk}(R)$
peaks at $R \simeq 4 R_S$, we can use the UV data to constrain
the peak of $T_{diks}(R)$ and we can use Eq. \ref{eq:T_disk} to
estimate $R_S$. From our data we get a value of $T_{disk}^{peak}\simeq 4\times 10^4$ K.
Assuming as a reference value for the accretion efficiency $\epsilon\simeq 0.1$
we get $R_S\simeq1.6\times10^{14}$ cm. The BH mass $M_{BH}$ is then 
about $5.4\times10^{8} M_{\odot}	$ and the accretion rate 
of about $0.5~ M_{\odot}$/year that corresponds to $\simeq 0.04$ times the
Eddington accretion rate ($\dot{M}_{Edd}$). The value of the BH mass estimated
from our UV data is compatible with that obtained by \cite{Oshlack2002}
using the virial assumption with measurements of the H$\beta$ FWHM and luminosity
($M_{BH}\simeq$ 3.9$\times10^8 M_{\odot}$), and by \cite{Xie2005}
($M_{BH}\simeq$ 2$\times10^8 M_{\odot}$).

The radius of the BLR can be estimated using the empirical relation by      \cite{Kaspi2005} and \cite{Bentz2006}:

\begin{equation}
R_{BLR}\simeq 20\Big[ \frac{\nu^* L(\nu^*)}
{10^{44}~ \textrm{erg}~\textrm{s}^{-1}} \Big]^{0.5}~~ \textrm{lt-days}.
\end{equation}
where $\nu^*\simeq 5.878\times10^{14}$ ~Hz.
According to the UVOT data we estimate $\nu^* L(\nu^*)<2\times10^{45}$ 
erg/s, and we get a value of $R_{BLR}< 2.3\times 10^{17}$ cm. 
We use BLR radius of about $1.6\times 10^{17}$ cm
with reflectivity value of 0.1 \citep{Ghisellini2009}. The DT 
is modelled as a BB with a temperature $T_{DT}$ of about 100 K,
and a reflectivity of 0.3 \citep{Ghisellini2009}. The distance is set to a typical
value of $\simeq 10^{18}$ cm, and it is fine tuned in the fit in order
that the BLR/DT+BLR/ERC correctly match the X-ray and hard X-ray data.

We use an emitting region size ($R_{rad}$) in the range $[2-5]\times10^{16}$ cm, that 
is almost twice the value estimated from the fast variability. In this regard, we note 
that since we are describing flare-averaged states, with integration times of the order of a few weeks, 
the discrepancy with the fast variability estimate is not problematic.  
As electron energy distribution $N(\gamma)$ we use a broken power-law:

\begin{equation}
N(\gamma)\propto\begin{cases}
 \gamma^{-p} ~~~~\mbox{for} ~~~\gamma_{min}\leq\gamma<\gamma_{br}\\
 \gamma^{-p1} ~~~\mbox{for} ~~~~~ \gamma_{br}\leq\gamma\leq\gamma_{max}
\end{cases}
\end{equation}

with the number density of the electrons: 

\begin{equation}
N_e=\int_{\gamma_{min}}^{\gamma_{max}} N(\gamma)d\gamma. 
\end{equation}
The low energy spectral index ($p$) is
chosen $\simeq 1.9$, as hinted by the typical X-ray photon index 
(see Sect. \ref{sec:gamma-ray-phys}); 
the value of the break energy ($\gamma_{br}$) is chosen to
match the position of the peak energy of the IC bump. It is
of the order of [250-300] and it is tuned in the three flares to
fit the data. The high energy spectral index ($p_1$) is
chosen according to the average spectral index of the
LAT data, and the reference value is $\simeq 3$. 
Magnetic field intensity $B$ is chosen to be 1.0 $G$ for flares
$b$ and $c$, while for the flare \textit{d} we need to use a larger value of \textit{B} 
and a more compact region size.

All the flares are assumed to occur at dissipation 
distance from the disk ($R_{diss}$) that is within the BLR and the DT
($R_{diss}<R_{BLR},~~ R_{diss}<R_{DT}$) .
The resulting best-fit of the MW SEDs are reported in Fig. 
\ref{fig:MWSED_BKN}, and the corresponding values of the best 
fit parameters for the three flares are reported in Tab. \ref{tab:SED-FIT-TAB-BKN}.

To have an indication of the change in the
energetic contents of the jet as a function of the flaring activity, we try to fit
also the quiescent \g-ray SED. 
Considering the lack of MW data during this time interval, we assume that the X-ray SED 
is close to the state observed during flare \textit{b}, as supported by the low X-ray variability.
From the MW light curves reported in Fig. \ref{fig:MWLC}, we 
infer that the optical/UV flux is at a level of few times $10^{-12}$ erg cm$^{-2}$ s$^{-1}$.
We can have an acceptable description of
the MW SED during the quiescent \g-ray period using parameters comparable
to those chosen for the flare $b$, but we need to decrease significantly
$\gamma_{br}$ to $\simeq 65$. The best-fit model is represented by the black dash-dotted line in 
the top panel of Fig. \ref{fig:MWSED_BKN}.

It is interesting to study the evolution of the energetic content of the jet resulting
from the SED modelling discussed above.
We evaluate the total kinetic power of the jet as:

\begin{equation}
\label{eq:L_jet}
L_{jet}=\pi R_{rad}^2 c \Gamma^2(U'_e + U'_B + U'_p)
\end{equation} 
where the rest frame magnetic energy density is given by $U'_B=B^2/8\pi$, 
the electron energy density is $U'_e$\footnote{$U'_e=\int_{\gamma_{min}}^{\gamma_{max}}\gamma m_ec^2 N(\gamma)d\gamma$}, 
and $U'_p=0.1N_{e}m_pc^2$ is the cold-proton energy density, assuming that we have one cold proton 
per ten electrons \citep{Sikora2009}. The power carried by the jet in terms of radiation is given by $L_{rad}\simeq L'\Gamma^2/4$. 
We evaluate $L'$ summing up the numerical integration of 
each radiative component (synchrotron, SSC, ERC/BLR, ERC/DT) as observed in the jet rest frame. 
In Tab. \ref{tab:SED-FIT-TAB-BKN-ENERGETIC}, we report the results for flare $b,c,d$ and the 
quiescent state, corresponding to the physical parameters reported in Tab. \ref{tab:SED-FIT-TAB-BKN}.

The total kinetic power of the jets is
almost steady, except during the flare $d$
when it increased by a factor of 2.
Indeed, for the case of flare $b$ and $c$, we find values
of $L_{jet}$ that are comparable with that of the quiescent state.
 On the contrary, we note that the value of the electron energy density during
all the flaring states is much larger than that estimated during the quiescent state,
and, during the flare $d$, it increased with respect to the quiescent state by about a factor of 2.7.\\
The low values of $U'_e/U'_B$ reflect the low SSC contribution.
Indeed, for the choice of our model parameters, the SSC is always negligible compared
to the other radiative components. 
As final remark, we note that the radiative efficiency of the jet ($L_{rad}/L_{jet}$), is 
$\simeq 0.03$ during the quiescent state, and increases up to $\simeq 0.2$ during 
flare $c$. 

We now compare the jet total kinetic luminosity with the accretion disk luminosity. 
According to the analysis reported in \cite{Ghisellini2009a}, if the jet power comes from 
the accretion process, then the accreting mass has to account both for the disk luminosity 
and for jet power. We can write:	

\begin{equation}
\epsilon_{tot}\dot{M}c^2=\epsilon_{D}\dot{M}c^2+\epsilon_{jet}\dot{M}c^2 .
\end{equation}
The ratio of $L_{jet}$ to $L_{D}$, reported in Tab. \ref{tab:SED-FIT-TAB-BKN-ENERGETIC} is equal
to $\epsilon_{jet}/\epsilon_{D}$. Using as $\epsilon_{D}$
the typical value of 0.1 we need a total efficiency $\epsilon_{tot} \gtrsim0.26$, 
that is still compatible with the maximum efficiency in the case of a Kerr BH.

\subsection{Comparison with other \textit{Fermi}-LAT FSRQs.}
It is interesting to compare the flaring activity of \PKS~with other FSRQs observed by \textit{Fermi}. 
3C~454.3 ($z$=0.859), reached an isotropic luminosity above 100 MeV 
$L(E>100)\simeq 8\times 10^{48}$ erg/s during the period of 2008 August/September  \citep{3C454Abdo2009}. 
It became the highest ever recorded object in the gamma rays during the flare in 2009 
December \citep{3C454.3ATEL2009}, with a luminosity $L(E>100)\simeq 1\times 10^{49}$ erg/s, about
five times larger than the maximum flare resolved luminosity of \PKS.  
The source during the flares exhibited rapid variability down to sub-daily timescales, that is comparable
to that observed in \PKS.
The \g-ray spectra of 3C~454.3 object show a spectral break 
around 2 GeV \citep{3C454Abdo2009}. In this case we see a relevant difference w.r.t. to \PKS,
that exhibited a mild curvature spectrum. A possible reason, as explained in the previous Section, is that 
 in the case of 3C~454.3 the dominant external photon field originates in the DT and 
 not in the BBB. Thus the IC process
occurs under the TH regime, and the break reflects the break in the electron distribution. 
The black hole mass is estimated to be $M_{BH}\simeq 	4\times 10^9 M_{\odot}$,
and the disk luminosity of the order of $L_D \simeq 2\times 10^{46}$ erg/s, both
these values are one order of magnitude larger than \PKS. 
Another interesting difference is that the 3C~454.3 exhibits
 no BBB features during the high optical and \g-ray outburst.

The FSRQ PKS 1502+106 ($z$=1.839), during the outburst in 2008 August  \citep{PKS1502_Abdo2009}, reached
a luminosity $L \simeq 1\times 10^{49}$ erg/s, showing
a very large Compton dominance up to 100. As in the case
of 3C~454.3, this object exhibited no BBB feature at UV energies. 
The flare of this object was an isolated episode, rather than
a flaring sequence. The \g-ray spectrum showed a curved shape,
 similar to \PKS, during the full integration period,
and during the post-flare state. The curvature values, 0.1-0.2,
are comparable to those observed in \PKS, hence also in this
case an origin in the KN effect is possible, as well as a curved
electron distribution with the IC process in TH regime.
The MW variability of this object is very different from that
observed in \PKS. Indeed, the X-ray flux followed both the \g-ray 
and the optical flux. Moreover, there was a strong correlation 
between the UV and the \g-ray. Probably this can be explained
with a sharp change in the density of the radiating electrons,
or in the beaming factor.

PKS 1454-354 ($z$=1.424), another distant FSRQs discovered in the \g-ray by \textit{Fermi}
during the 2008 August/September  flaring activity \citep{PKS1454Abod2009},
reached a luminosity $L(E>100) \simeq 7\times 10^{48}$ erg/s, and exhibited a rapid variability with
 hours/days timescales.
 
We conclude that the most rapid timescale for the \g-ray variability (down to 6/12 hr) is common
to this class of objects, despite the bolometric \g-ray luminosity. 
This timescale is limited by the minimum LAT
integration time required to extract a flux for object with $F(E>200 MeV)$ of
the order of $10^{-5}~$\latflux, hence a faster variability could be possible.
Compared to those listed above, \PKS~seems
to be a less powerful object in terms of both \g-ray luminosity and
thermal luminosity.
The curvature in the \g-ray spectrum, when compared to the spectral break
observed in other sources such as 3C~454.3, may be understood as a signature of the KN regime, 
hence of a dissipation region within the BLR.

\section{Conclusions} 
We presented MW observations of \PKS~during a period of about 11 months,
when the source exhibited a strong evolution of its broad-band SED,
characterized by a complex variability both at Optical/UV and \g-ray energies,
with timescales detected down to the level of 6/12 hr. 
The \g-ray flux shows the usual harder when brighter trend,
for flux levels of $F(E>0.2 GeV) \gtrsim $ 2.4\latfluxvii. For lower flux levels
the trend seems to be the opposite, but the low statistics during
these dimmer states does not allow us to give a purely physical
interpretation. The spectrum shows a mild curvature, both during the flares
and during the full-period integration, well described
by a log-parabolic law, and can be understood as a signature of the KN effect.

The \g-ray flux shows a complex correlation with the other
wavelengths. There is no correlation at all with the X-ray band,
a week correlation with the UV, and a strong correlation with the
optical ($R$) flux. 
{ Thanks to the unprecedented continuous LAT \g-ray sky survey, 
we were able to find a lag in the \g-ray light curve, with the \g-ray band leading the $R$-band by 
about 13 days}. Because of this complex multi-band variability,
we assume that a change in the beaming factor cannot account
for the flaring activity of this object. Indeed, assuming as the
main driver a change in the beaming factor, we would expect
at least a weak correlation between the X-ray and the optical band,
or between the X-ray and the \g-ray. This is also in agreement
with the absence of correlation between the jet kinematic
power($L_{jet}$) and the flaring episodes observed in our
data set. 

UV data allowed us to estimate the mass of the BH of 
$\simeq 5.6\times 10^8 M_{\odot}$. This value, that is in agreement
with other estimates based on different methods,
is smaller compared
to the very luminous and distant FSRQs, with BH masses of the
order of $10^9 M_{\odot}$. As a consequence of the estimated
BH mass and 
thermal component luminosity, also the accretion 
rate of $\simeq 0.04 \dot{M}_{Edd}$, is lower when
compared to the expectation in case of FSRQs ($\simeq 0.1 \dot{M}_{Edd}$).\\
Due to the low redshift of the source, the bolometric isotropic \g-ray luminosity is also smaller
compared to other distant FSRQs observed by \textit{Fermi}. Indeed,
\PKS~has a typical \g-ray
luminosity and BH mass about one order of magnitude lower
 compared to sources like 3C~454.3 or PKS 1502-106.

Despite the relatively lower power in the thermal
and non-thermal outputs, \PKS~exhibits a quite large
Compton dominance, as observed
in the most powerful FSRQs, and prominent a BBB signature.
The object could be a representative of an aged
FSRQ, hence the analysis here presented is relevant
in order to understand the evolution of these
objects. 


We note the puzzling feature of the BBB UV shape. 
The BBB was still prominent
during the historical maximum optical state during 2009 
May, although the optical/UV spectral index
was softer compared to that in quiescent state. 

The analysis presented here shows the importance 
of the MW monitoring of blazars, independent of the \g-ray 
triggering. Indeed, only comparing the flaring
and quiescent states, and understanding the evolution
of the physical parameters as a function of the flaring
activity, is it possible to discriminate among the
possible physical scenarios.

{\it Acknowledgments}\\
The \latf~ Collaboration acknowledges the generous
support of a number of agencies and institutes that have supported the $Fermi$
LAT Collaboration. These include the National Aeronautics and Space
Administration and the Department of Energy in the United States, the
Commissariat \`a l'Energie Atomique and the Centre National de la Recherche
Scientifique / Institut National de Physique Nucl\'eaire et de Physique des
Particules in France, the Agenzia Spaziale Italiana and the Istituto Nazionale
di Fisica Nucleare in Italy, the Ministry of Education, Culture, Sports,
Science and Technology (MEXT), High Energy Accelerator Research Organization
(KEK) and Japan Aerospace Exploration Agency (JAXA) in Japan, and the K.\ A.\
Wallenberg Foundation, the Swedish Research Council and the Swedish National
Space Board in Sweden. {Additional support for science analysis during the operations 
phase is gratefully acknowledged from the Istituto Nazionale di Astrofisica in Italy 
and the Centre National d'\'Etudes Spatiales in France.}\\
Part of this work is based on archival data and on bibliographic information 
obtained from the NASA/IPAC Extragalactic  Database (NED).\\ 
{This paper is partly based on observations carried out at the German-Spanish 
Calar Alto Observatory, which is jointly operated by the MPIA and the IAA-CSIC.\\}
{Acquisition of the MAPCAT data is supported in part by the Spanish Ministry of Science and 
Innovation and the Regional Government of Andaluc\'{\i}a through grants AYA2007-67626-C03-03 and P09-FQM-4784, respectively.\\}
This work was partly supported by the Italian Space Agency through 
contract ASI-INAF I/088/06/0 for the Study of High-Energy Astrophysics.
St.Petersburg University team acknowledges support from Russian RFBR foundation via grant 09-02-00092.
AZT-24 observations are made within an agreement between  Pulkovo,
Rome and Teramo observatories. \\
Abastumani Observatory team acknowledges financial support by the
Georgian National Science Foundation through grant GNSF/ST08/4-404.\\
The research at Boston University (BU) was funded in part by NASA Fermi Guest Investigator grants NNX08AV65G and 
NNX08AV61G,  and by the National Science Foundation (NSF) through grant AST-0907893.The PRISM camera at the Perkins Telescope of
Lowell Observatory was developed by K.\ Janes et al. at BU and Lowell Observatory, with funding from the NSF, BU, and Lowell Observatory.
The Liverpool Telescope is operated on the island of La Palma by Liverpool John Moores University in the
Spanish Observatorio del Roque de los Muchachos of the Instituto de Astrofisica de Canarias, with funding
from the UK Science and Technology Facilities Council.\\
This research has made use of observations from the MOJAVE 
database that is maintained by the MOJAVE team (Lister et al. 
2009b). The MOJAVE project is supported under National Science 
Foundation grant 0807860-AST and NASA-Fermi grant NNX08AV67G.\\
The Lebedev Physical Institute team was partly supported by the 
Russian Foundation for Basic Research (project 08-02-00545).
KS was supported by stipend from the IMPRS for Astronomy and Astrophysics.
We thank A. P. Marscher and S. Jorstad for providing multiwavelength data.
{\it Facilities:} {Fermi LAT, Swift XRT, UVOT, GASP-WEBT,NRAO(VLBA)}. 
{The VLBA is a facility of the National 
Science Foundation operated by the National Radio Astronomy 
Observatory under cooperative agreement with Associated 
Universities, Inc.}

\bibliography{PKS1510-089.bib} 
 
\newpage 
\begin{table}[!htp]
\caption{Flaring activity of \PKS~ from August 2008 until June 2009}
\begin{flushleft} 
 \begin{tabular}{lllll} 
 \hline
 source state& date start & date stop & MJD start & MJD stop\\
 \hline
 flare \textit{a} & 2008-08-30 & 2008-09-26 &54708  &54735\\
 quiescent      &  2009-09-30 & 2010-01-01 &54739  &54832\\
 flare \textit{b} & 2009-01-04 & 2009-01-27 &54835  &54858\\
 flare \textit{c} & 2009-03-10 & 2009-04-09 &54900  &54930\\
 flare \textit{d} & 2009-04-15 & 2009-05-12 &54936  &54963\\
 \hline
	\end{tabular}
\label{tab:flares}
\end{flushleft} 
\end{table}
\begin{table}[!htp]
 \caption{Unbinned likelihood LRT summary }
   \begin{flushleft} 
{\small
  \begin{tabular}{lllllll}
 \hline
   time range& -LogLike(PL)& -LogLike(LP)&-LogLike(BPL)&LRT(PL/LP)/P.(LP)$^{**}$ &LRT(LP/BPL)/P.(BPL)$^{**}$\\
   \hline
   \hline
   full            & 323082.6     & 323056.4      & 323062.8          & 56.2/$<10^{-5}\%$      & -12.8/NULL   \\
                   &              &               & 323059.0$^{*}$	  &                        & -5.2/NULL \\
 \hline   
   quiescent       & 83057.7      & 83057.5       & 83057.6           & 0.4/$47.3\%$           & -0.2/NULL   \\
 \hline  
   flare a         & 28908.9      & 28902.4       & 28903.5           & 6.5/$99.97\%$          & -2.2/NULL   \\
  				   &              &               & 28902.5$^{*}$	  &                        & -2.0/NULL  \\
 \hline			   
   flare b         & 23932.7      & 23928.5       & 23927.9           & 8.4/$99.62\%$          & 1.2/$72.67\%$         \\ 
                   &              &               & 23927.9$^{*}$	  &                        & 1.2/$72.67\%$ \\
  \hline
   flare c         & 38328.9      & 38318.5       & 38320.2           & 20.8/$99.99\%$         & -3.4/NULL         \\  
                   &              &               & 39319.4$^{*}$	  &                        & -1.8/NULL  \\ 
  \hline   
   flare d         & 31326.1      & 31322.0       & 31323.3           & 8.2/$99.58\%$          &-2.6/NULL                 \\ 
                   &              &               & 31322.8$^{*}$	  &                        &-1.6/NULL  \\
   \hline 
    \hline 
  	
  \end{tabular}
\label{tab:LAT_LRT}
$^{*}$ BPL fit by means of loglikelihood profile\\
{$^{**}$ P.(LP) and P.(BPL) are the cumulative distribution functions of the LRT statistics, evaluated at
the LRT value actually observed. These probabilities are evaluated using as reference distribution  a   
$\chi^2_d$ distribution with the number of degree of freedom ($d$) equal to the  difference in the number of 
free parameters in the two models.}
}
\end{flushleft} 
\end{table}
\begin{sidewaystable}[!h]
 \caption{Unbinned likelihood spectral fit results: }
   \begin{flushleft} 
{\tiny
  \begin{tabular}{lccc|cccc|ccccc}
 \hline
                    &  \multicolumn{3}{c}{PL}                  &   \multicolumn{4}{c}{LP}                                    &   \multicolumn{5}{c}{BPL$^{***}$} \\
 time range& $\alpha_{\gamma}$& $F_{100}^{*}$& -LogLike &$\alpha_{\gamma}$ &$\beta$  &$F_{100}^{*}$  &-LogLike &$\alpha_{\gamma}$ &$\alpha_{\gamma_1}$ &$E_b^{**}$& $F_{100}^{*}$  &-LogLike\\
                       \hline
  \hline
full              & 2.44$\pm$0.01 & 1.32$\pm$0.03 & 323082.6& 2.23$\pm$0.02 & 0.09 $\pm$ 0.01& 1.12$\pm$0.03& 323056.4 &2.30$\pm$0.02  &2.63 $\pm$ 0.04 &980$\pm$130  &1.37$\pm$01    & 83057.6\\       
                   &                             &                              &                  &                             &                               &                             &                  &------                      &------                       &------------           &--------                &-------------\\
quiescent &2.43$\pm$0.07 & 0.27$\pm$0.03 & 83057.7  & 2.3$\pm$0.1      & 0.03 $\pm$ 0.05& 0.25$\pm$0.04& 83057.5    &2.5$\pm$1.5      &2.4 $\pm$ 1.5      &700$\pm$8000 &0.3$\pm$02     & 28903.5\\
                  &                             &                              &                  &                             &                               &                             &                  &------                      &------                       &------------           &--------                &-------------\\
flare a       & 2.55$\pm$0.01& 1.55$\pm$0.02 & 28908.9   & 2.2$\pm$0.1      & 0.19 $\pm$ 0.06& 1.16$\pm$0.12& 28902.4    &2.39$\pm$0.01 &3.22 $\pm$ 0.04 &1500$\pm$40   &1.37$\pm$0.01 & 28903.5\\
			 &                             &                              &                  &                             &                               &                             &                  &------                      &------                       &------------           &--------                &-------------\\
flare b       & 2.35$\pm$0.04 & 2.1  $\pm$0.1   & 23932.7   & 2.0$\pm$0.1      & 0.10 $\pm$ 0.03&  1.7$\pm$0.1    & 23928.5   &2.26$\pm$0.05 &3. $\pm$0.03      &3400 $\pm$800& 1.9$\pm$0.1    &  23927.9  \\
                  &                             &                              &                  &                             &                               &                             &                  &------                      &------                       &------------           &--------                &-------------\\
flare c       & 2.37$\pm$0.3 & 3.9  $\pm$0.1   & 38328.9   & 2.13$\pm0.06$  &0.1  $\pm$ 0.02&  3.3$\pm$0.2    & 38318.5   &2.28$\pm$0.04 &2.9$\pm$0.2    &1918 $\pm$593   &3.7 $\pm$0.1    &  38319.9 \\
                   &                             &                              &                  &                             &                               &                             &                  &------                      &------                       &------------           &--------                &-------------\\
flare d       & 2.44 $\pm$0.04& 2.9  $\pm$0.1   &31326.1   & 2.24$\pm$0.08   &0.09  $\pm$ 0.03&  2.5$\pm$0.1    & 31322.0   &2.31$\pm$0.08 &2.6 $\pm$0.1     &1000 $\pm$300  &2.6 $\pm$0.1  & 31323.3 \\
  \end{tabular}
\label{tab:LAT_FIT}
\\
$^{*}$ $10^{-6}$ ph cm$^{-2}$ s$^{-1}$\\
$^{**}$ MeV\\
$^{***}$  BLP fit by means of loglikelihood profile\\
}
\end{flushleft} 
\end{sidewaystable}

     \begin{table*}
    \caption{Spectral analysis of XRT data. {In the last column we report the reduced $\chi^2$ and, in parentheses, the degree of freedom.
}}
     \label{tab:xrt_fit} 
     \begin{flushleft} 
     {\small
    \begin{tabular}{cccccc}
    \hline 
     \hline 
     Observation Date &Start Time   &Norm  &$\alpha_{X}$ &flux 0.3-10 keV                    &$\chi^2_r$/dof \\ 
                      &MJD (days)   &$10^{-4}$        &          &$10^{-12} erg~ cm^{-2} ~s^{-1}$  & \\ 
    \hline 
    \noalign{\smallskip}
2009-01-10  &54841.8  &9.4$^{+0.6}_{-0.6}$ &1.35$^{+0.07}_{-0.08}$  &8.5$^{+0.8}_{-0.6}$   &0.697(43) \\ 
2009-01-11  &54842.9  &9.4$^{+0.6}_{-0.6}$ &1.28$^{+0.07}_{-0.07}$  &9.4$^{+0.7}_{-0.7}$   &0.817(46) \\ 
2009-01-13  &54844.0  &9.9$^{+0.7}_{-0.7}$ &1.46$^{+0.08}_{-0.08}$  &7.9$^{+0.6}_{-0.6}$   &0.978(38) \\ 
2009-01-14  &54845.8  &10.4$^{+0.7}_{-0.6}$ &1.41$^{+0.07}_{-0.07}$  &8.8$^{+0.7}_{-0.9}$   &0.811(46) \\ 
2009-01-16  &54847.7  &10.2$^{+0.7}_{-0.7}$ &1.45$^{+0.07}_{-0.07}$  &8.2$^{+0.8}_{-0.8}$   &0.778(42) \\ 
2009-01-16  &54849.2  &8.0$^{+0.8}_{-0.8}$ &1.44$^{+0.12}_{-0.12}$  &6.5$^{+0.8}_{-0.8}$   &0.571(20) \\ 
2009-01-21  &54852.8  &9.8$^{+0.6}_{-0.6}$ &1.41$^{+0.07}_{-0.07}$  &8.3$^{+0.7}_{-0.7}$   &0.898(51) \\ 
2009-01-25  &54856.8  &9.6$^{+0.7}_{-0.7}$ &1.36$^{+0.09}_{-0.09}$  &8.7$^{+0.8}_{-0.8}$   &0.832(30) \\ 
2009-03-06  &54896.9  &8.9$^{+1.2}_{-1.2}$ &1.5$^{+0.2}_{-0.2}$  &7.1$^{+1.2}_{-1.4}$   &0.980(11) \\ 
2009-03-11  &54901.6  &11.8$^{+0.6}_{-0.6}$ &1.62$^{+0.06}_{-0.06}$  &7.8$^{+0.6}_{-0.5}$   &1.067(68) \\ 
2009-03-12  &54902.6  &13.1$^{+0.6}_{-0.6}$ &1.58$^{+0.05}_{-0.06}$  &9.1$^{+0.6}_{-0.6}$   &0.897(78) \\ 
2009-03-17  &54907.2  &10.7$^{+0.6}_{-0.6}$ &1.6$^{+0.07}_{-0.07}$  &7.2$^{+0.6}_{-0.6}$   &1.184(61) \\ 
2009-03-18  &54908.0  &9.9$^{+0.5}_{-0.5}$ &1.45$^{+0.06}_{-0.06}$  &8$^{+0.6}_{-0.5}$   &0.949(59) \\ 
2009-03-19  &54909.8  &12.1$^{+0.8}_{-0.8}$ &1.69$^{+0.09}_{-0.09}$  &7.4$^{+0.8}_{-0.6}$   &1.049(35) \\ 
2009-03-20  &54910.9  &8.7$^{+1}_{-1}$ &1.3$^{+0.14}_{-0.14}$  &8.5$^{+1.2}_{-1.2}$   &1.210(20) \\ 
2009-03-22  &54912.1  &9.9$^{+0.9}_{-0.8}$ &1.46$^{+0.11}_{-0.11}$  &7.9$^{+0.8}_{-0.8}$   &1.325(30) \\ 
2009-03-22  &54912.1  &16.5$^{+1.4}_{-1.4}$ &1.69$^{+0.1}_{-0.1}$  &10.1$^{+1.2}_{-1.2}$   &0.506(20) \\ 
2009-03-23  &54913.5  &9.3$^{+0.7}_{-0.7}$ &1.47$^{+0.1}_{-0.1}$  &7.3$^{+0.6}_{-0.7}$   &0.983(30) \\ 
2009-03-24  &54914.1  &10.2$^{+0.9}_{-0.9}$ &1.56$^{+0.12}_{-0.11}$  &7.3$^{+0.8}_{-0.7}$   &0.770(23) \\ 
2009-03-25  &54915.6  &11.1$^{+0.8}_{-0.8}$ &1.51$^{+0.08}_{-0.08}$  &8.4$^{+0.9}_{-1}$   &1.293(37) \\ 
2009-03-26  &54916.7  &10.6$^{+1}_{-1}$ &1.54$^{+0.12}_{-0.12}$  &7.6$^{+1.1}_{-1.3}$   &1.084(19) \\ 
2009-03-27  &54917.2  &9.9$^{+1}_{-1}$ &1.53$^{+0.12}_{-0.12}$  &7.3$^{+1.1}_{-0.8}$   &0.840(19) \\ 
2009-03-28  &54918.2  &10.0$^{+0.7}_{-0.7}$ &1.35$^{+0.08}_{-0.08}$  &9.1$^{+0.9}_{-0.7}$   &0.777(32) \\ 
2009-03-30  &54920.4  &11.8$^{+0.9}_{-0.9}$ &1.41$^{+0.08}_{-0.08}$  &10$^{+1}_{-1}$   &0.855(31) \\ 
2009-04-04  &54925.5  &8.1$^{+1.3}_{-1.4}$ &1.2$^{+0.2}_{-0.2}$  &8$^{+1}_{-2}$   &1.775(12) \\ 
2009-04-10  &54931.0  &10.5$^{+0.7}_{-0.6}$ &1.5$^{+0.07}_{-0.07}$  &8$^{+0.6}_{-0.7}$   &1.098(46) \\ 
2009-04-27  &54948.6  &12.7$^{+1.3}_{-1.3}$ &1.62$^{+0.13}_{-0.13}$  &8.4$^{+1}_{-1.3}$   &0.502(15) \\ 
2009-04-28  &54949.6  &12.0$^{+2}_{-2}$ &1.13$^{+0.13}_{-0.13}$  &15$^{+3}_{-2}$   &0.446(13) \\ 
2009-04-29  &54950.8  &11.7$^{+1.1}_{-1}$ &1.36$^{+0.11}_{-0.11}$  &10.5$^{+1.2}_{-1.1}$   &0.951(25) \\ 
2009-05-01  &54952.6  &10.0$^{+2}_{-2}$ &1.5$^{+0.2}_{-0.2}$  &7.7$^{+1.7}_{-1.2}$   &1.327(08) \\ 
2009-05-02  &54953.4  &13.0$^{+1}_{-1}$ &1.58$^{+0.1}_{-0.1}$  &9$^{+1.2}_{-0.8}$   &1.241(29) \\ 
2009-05-03  &54954.8  &11.9$^{+1.5}_{-1.5}$ &1.5$^{+0.2}_{-0.2}$  &9$^{+2}_{-2}$   &0.838(10) \\ 
2009-05-04  &54955.7  &10.0$^{+2}_{-2}$ &1.2$^{+0.2}_{-0.2}$  &11$^{+2}_{-3}$   &1.428(08) \\ 
2009-05-05  &54956.7  &10.5$^{+0.9}_{-0.9}$ &1.53$^{+0.11}_{-0.12}$  &7.7$^{+0.7}_{-0.8}$   &0.838(22) \\ 
2009-05-07  &54958.4  &10.1$^{+0.8}_{-0.8}$ &1.28$^{+0.08}_{-0.08}$  &10.1$^{+0.9}_{-1}$   &1.122(30) \\ 
2009-05-11  &54962.5  &5.4$^{+1.4}_{-1.5}$ &0.9$^{+0.3}_{-0.3}$  &10$^{+3}_{-2}$   &1.660(05) \\ 
2009-05-12  &54963.3  &13.6$^{+0.7}_{-0.7}$ &1.43$^{+0.06}_{-0.06}$  &11.2$^{+0.8}_{-0.8}$   &0.931(62) \\ 
2009-05-13  &54964.1  &12.3$^{+0.6}_{-0.6}$ &1.6$^{+0.06}_{-0.06}$  &8.3$^{+0.5}_{-0.6}$   &0.822(59) \\ 
2009-05-14  &54965.3  &11.1$^{+0.6}_{-0.6}$ &1.48$^{+0.06}_{-0.06}$  &8.6$^{+0.6}_{-0.8}$   &1.175(56) \\ 
2009-05-30  &54981.4  &12.0$^{+4}_{-4}$ &1.7$^{+0.4}_{-0.5}$  &7$^{+3}_{-2}$   &0.768(02) \\ 
2009-06-07  &54989.7  &17.0$^{+2}_{-2}$ &1.8$^{+0.2}_{-0.2}$  &9.2$^{+1.2}_{-1.4}$   &1.575(16) \\ 
2009-06-13  &54995.6  &11.6$^{+1.3}_{-1.3}$ &1.35$^{+0.14}_{-0.14}$  &10.6$^{+1.5}_{-1.2}$   &1.185(20) \\ 
2009-06-20  &55002.5  &16.0$^{+1.2}_{-1.2}$ &1.86$^{+0.1}_{-0.1}$  &8.3$^{+0.8}_{-0.8}$   &0.888(24) \\ 
\hline
    \noalign{\smallskip}
    \end{tabular}
    }
    
    \end{flushleft}

\end{table*}

 \begin{table*}
    \caption{Best-fit parameters for the SED modelling in the case of BPL electron distribution.}
     \label{tab:SED-FIT-TAB-BKN} 
     \begin{flushleft} 
  {\small  
    \begin{tabular}{lllllllllllllll}

    \hline 
    \hline	
    
    flare        &$L_D$ &$M_{BH}$    &$R_{BLR}$ &$R_{DT}$ &$\Gamma$  &$R_{rad}$ &$B$ &$N_e$  &$\gamma_{min}$ &$\gamma_{max}$ &$\gamma_{br}$ &$p$ &$p_1$ \\
    	        
 	             &{\tiny $10^{45}$ erg/s} &{\tiny $10^8M_{\odot}$}  &{\tiny$10^{17}$cm}  &{\tiny$10^{17}$cm}        &          &{\tiny $10^{17}$ cm}  &{\tiny G}   &{\tiny \#/cm$^3$} &                &               &              &    & \\		   
	   
    \hline
	\textit{b}   &3.0 &5.6 &1.6  &12.0   &14.0 &0.32  &1.0 &550   &1.0 &2.2$\times10^4$ &220 &1.95 &3.15\\
	\textit{c}   &--  &--  &--   &--     &15.5 &0.32  &1.1 &300   &1.0 &2.2$\times10^4$ &280 &1.80 &3.15\\
	\textit{d}   &--  &--  &--   &--     &16.0 &0.25  &2.2 &800   &1.0 &7.0$\times10^3$ &200 &1.90 &3.20\\
	quiescent    &--  &--  &--   &--     &12.0 &0.43  &1.0 &350   &1.0 &2.2$\times10^4$ &65  &1.95 &3.20\\ 
    \hline
    \noalign{\smallskip}   
    \end{tabular}
	$R_{diss}<R_{BLR},~~ R_{diss}<R_{DT},~~ T^{peak}_{disk}=4\times 10^4 K,~~ T_{DT}=100 K,~~\theta=2.5^{\circ} $   
    }
    \end{flushleft} 
    
 \end{table*}

%
%
%
%
%
%
%
%
%
%

  \begin{table*}
    \caption{Jet power for the case of BPL modellig.}
     \label{tab:SED-FIT-TAB-BKN-ENERGETIC} 
     \begin{flushleft} 
  {\small  
    \begin{tabular}{lllllllll}

    \hline 
    \hline	
    
    flare   &$L_{jet}$ &$L_{rad}$ &$L_{e}/L_{jet}$ &$L_{B}/L_{jet} $&$L_{rad}/L_{jet}$ &$U'_e$ &$U'_e/U'_B$& $L_{jet}/L_{D}$   \\
    	        
 	        &{\tiny $10^{45}$ erg/s} &{\tiny $10^{45}$ erg/s} &&&&{\tiny erg/cm$^3$}&\\		   
	   
    \hline
	\textit{b}  &2.32  &0.26 &0.025 &0.316 &0.112  &0.0031 &0.08 &0.77  \\
	\textit{c}  &2.17  &0.46 &0.027 &0.502 &0.211  &0.0026 &0.05 &0.72  \\
	\textit{d}  &4.83  &0.34 &0.016 &0.606 &0.078  &0.0049 &0.03 &1.61  \\
	quiescent   &2.31  &0.07 &0.016 &0.423 &0.031  &0.0015 &0.04 &0.77  \\
	
    \hline
    \noalign{\smallskip}   
    \end{tabular}
	 
    }
    \end{flushleft} 
    
 \end{table*}



\begin{minipage}[t]{1\linewidth}

\begin{figure}[H]
\begin{center}
\epsfig{file=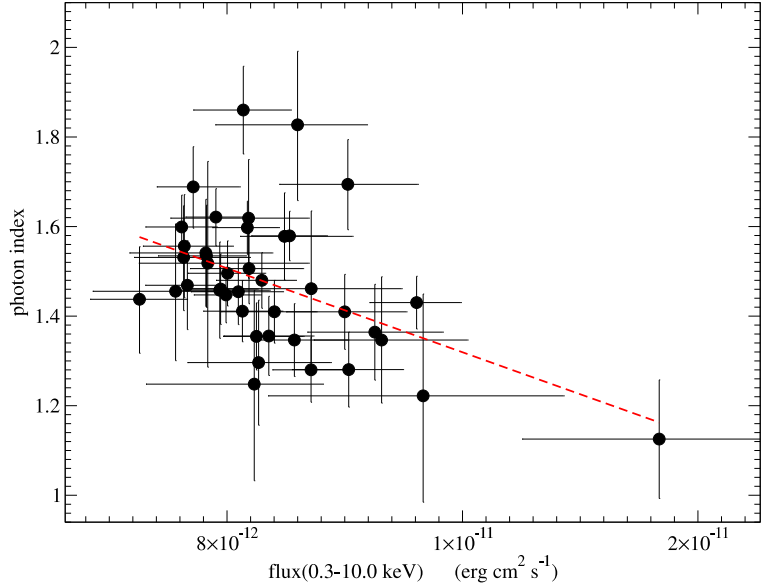,width=10cm,angle=0}
\end{center}
\caption{Scatter plot of the flux in $0.3-10.0$ keV range vs. the X-ray photon 
index ($\alpha_{X}$). The dashed lines represent a linear bet fit model.
Using the Monte Carlo method described in Sect. \ref{sec:Fermi-Spec-Ev},
we get a correlation coefficient $r= -0.31$ with a $95\%$ confidence
limit of $-0.55\leq r \leq-0.05 $}
\label{fig:XRT-FULX_VS_INDEX}
\end{figure} 

\begin{figure}[H]
\begin{center}
\epsfig{file=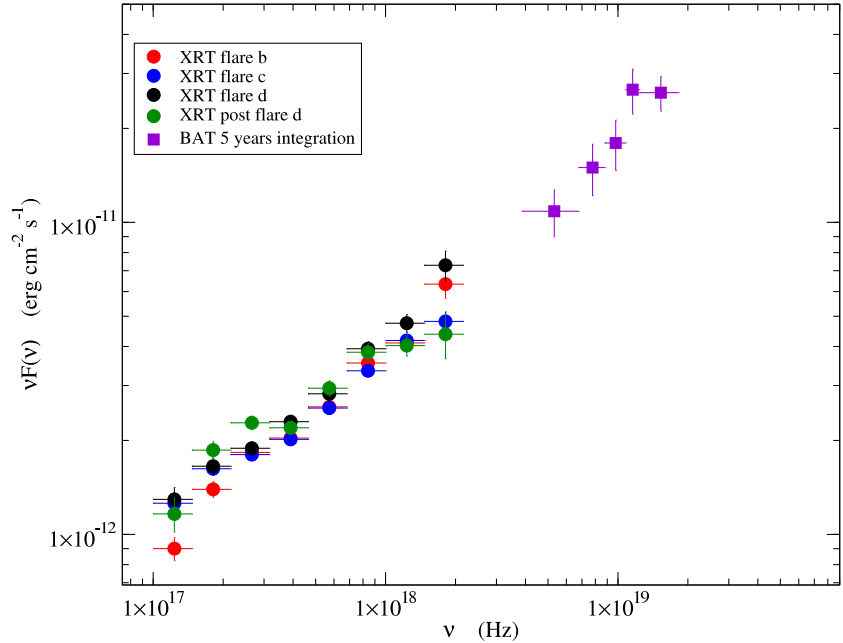,width=11cm,angle=0}
\end{center}
\caption{XRT SEDs averaged during the flare \textit{b}, \textit{c}, \textit{d} 
and duirng the post \textit{d} flare period. The violet boxes represent
the BAT spectrum averaged over 5 years} \label{fig:XRT-SEDs_AVERAGED} 
\end{figure} 
\end{minipage}

\begin{minipage}[t]{1\linewidth}
\begin{figure}[H]
\begin{center}
\epsfig{file=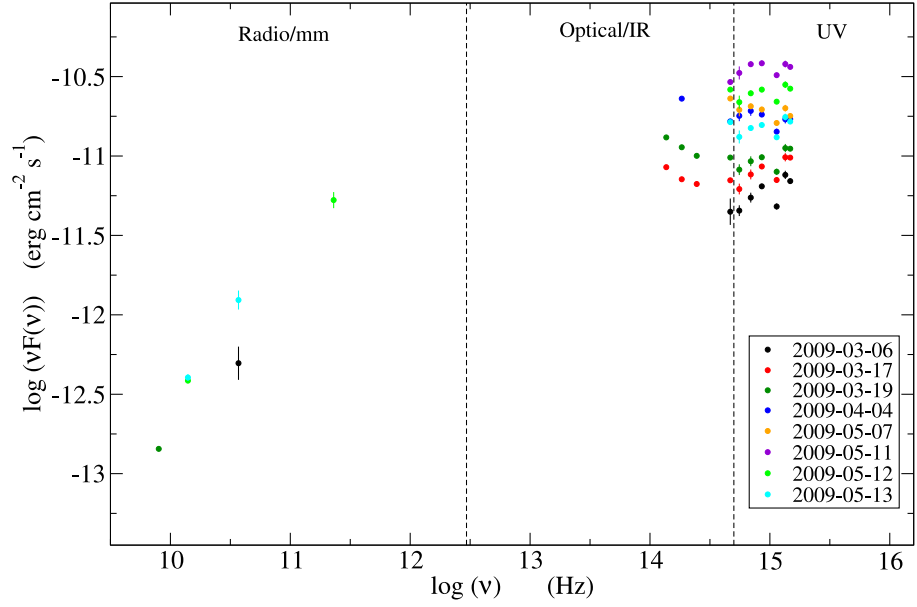,angle=0,width=12cm}
\end{center}
\caption{Radio to UV SEDs built using only data simultaneous within
a daily time span. }
\label{fig:UVOT_GASP_SED}
\end{figure}

\begin{figure}[H]
\begin{center}
\epsfig{file=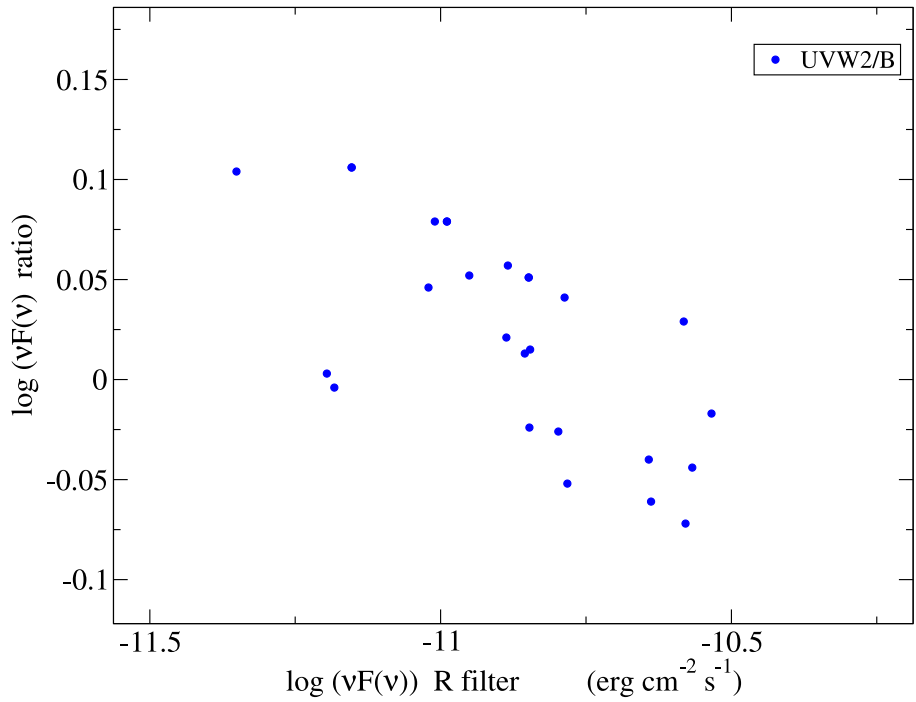,width=12cm}
\end{center}
\caption{The ratio of $\nu$F($\nu$)  in the UVOT UVW2 filter to $\nu$F($\nu$) 
in the UVOT \textit{B} filter, as a function of $\nu$F($\nu$) in the optical \textit{R} filter.
}
\label{fig:UVOT_GASP_SPEC_EVOL}
\end{figure}
\end{minipage}

\begin{minipage}[t]{1\linewidth}
\begin{figure}[H]
\begin{center}
\epsfig{file=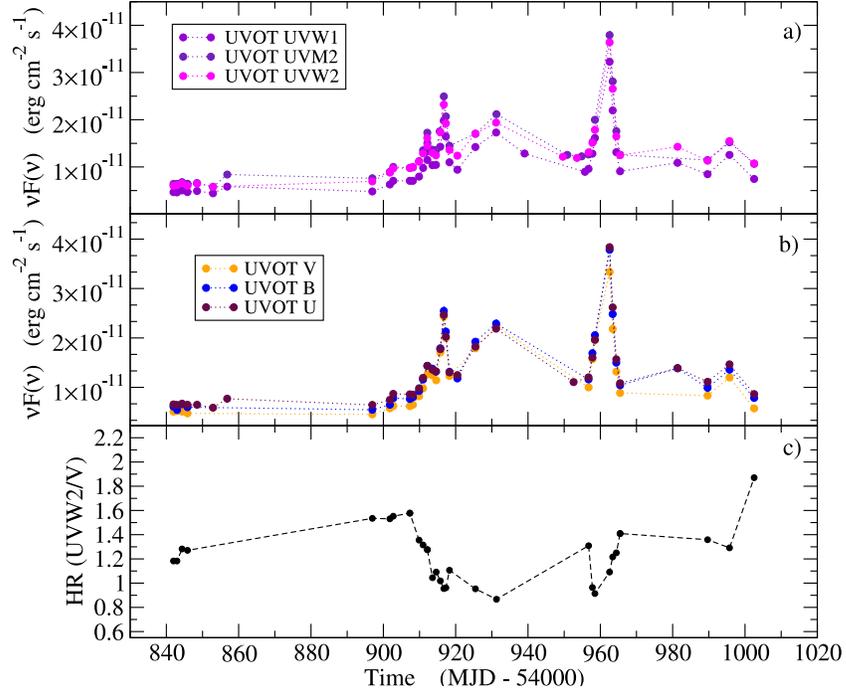,width=11cm,angle=0}
\end{center}
\caption{{\it Panels a,b:} UVOT light curves. {\it Panel c:} Hardness ratio of the
UVOT spectra (UVW2/UV\textit{V}) evaluated as a function of the time. }
\label{fig:UVOT_LC}
\end{figure}

\begin{figure}[H]
\begin{center}
\epsfig{file=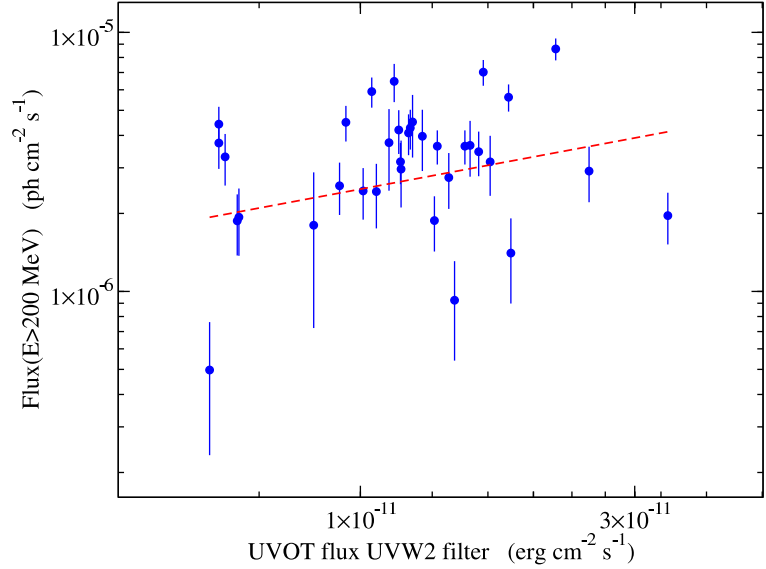,width=10cm,angle=0}
\end{center}
\caption{Scatter plot of the \textit{Fermi}-LAT flux (E$>200$ MeV) vs. the UVOT 
$\nu$F($\nu$) in the UVW2 filter. The red dashed line represents the best-fit 
power law model. The correlation coefficient of the logarithm of the UV and 
$\gamma$-ray fluxes is $r=0.2$ 
with a $95\%$ confidence interval of $0.05 \leq r \leq 0.34$.
} 
\label{fig:LAT-UVOT-CORR} 
\end{figure} 
\end{minipage}

\begin{minipage}[t]{1\linewidth}
\begin{figure}[H]
\begin{center}
\epsfig{file=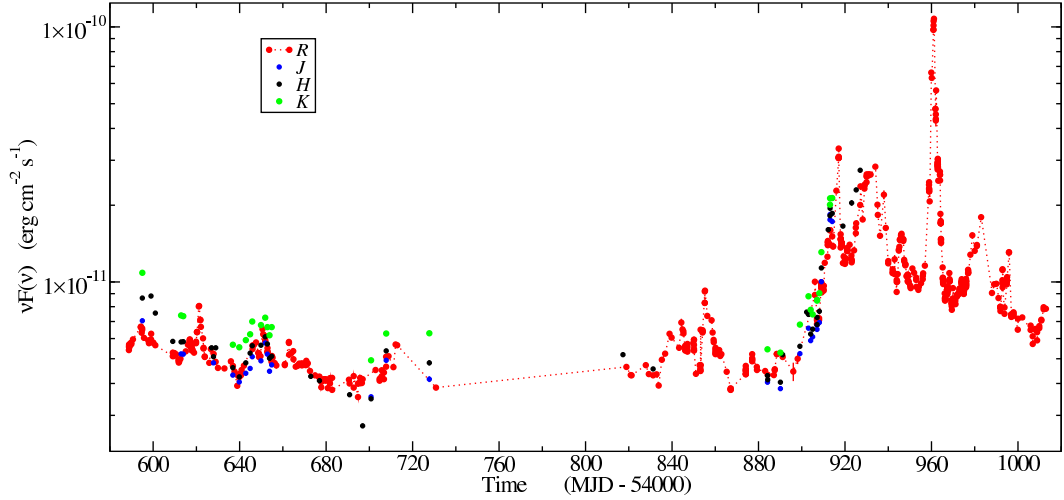,width=14cm,angle=0}
\end{center}
\caption{ Optical and near-IR light curves from the GASP project.}
\label{fig:GASP_OPT_LC}
\end{figure}

\begin{figure}[H]
\begin{center}
\epsfig{file=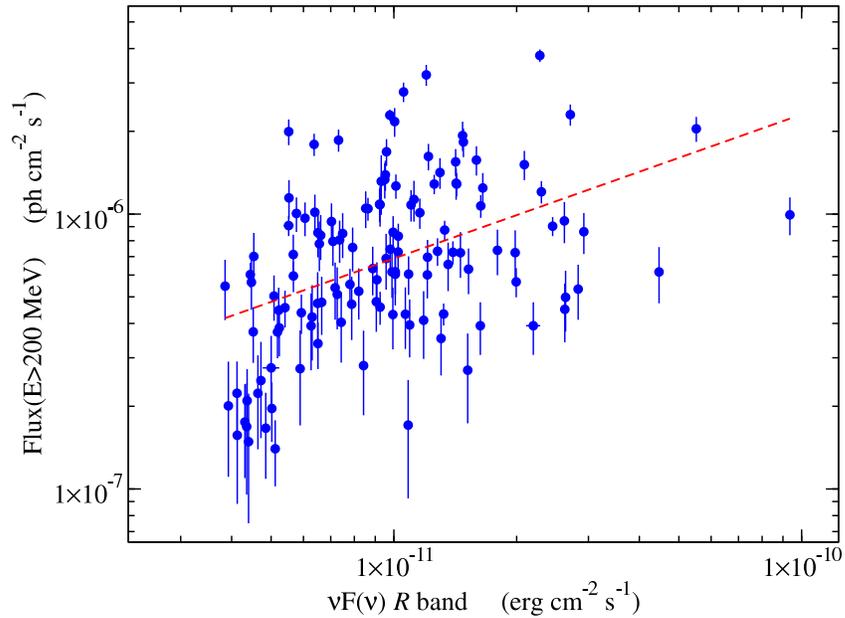,width=11cm,angle=0}
\end{center}
\caption{Scatter plot of the \textit{Fermi}-LAT flux(E$>200$ MeV) vs the $\nu F(\nu)$ in 
the optical R band. The dashed line represents the best fit power-law model. 
The correlation coefficient of the logarithm of the 
optical and $\gamma$-ray fluxes, evaluated through the Monte Carlo
method described in Sect. \ref{sec:Fermi-Spec-Ev}, is $r= 0.42$ 
with a $95\%$ confidence interval $0.36\leq r\leq 0.46$. 
There is an evident dispersion in the scatter 
plot, probably due to the inter-band time lags showed in Sect. 
\ref{sec:Optical and UV results} } 
\label{fig:LAT-OptR-CORR} 
\end{figure} 
\end{minipage}

\begin{minipage}[t]{1\linewidth}
\begin{figure}[H]
\begin{center}
\includegraphics[width=14cm]{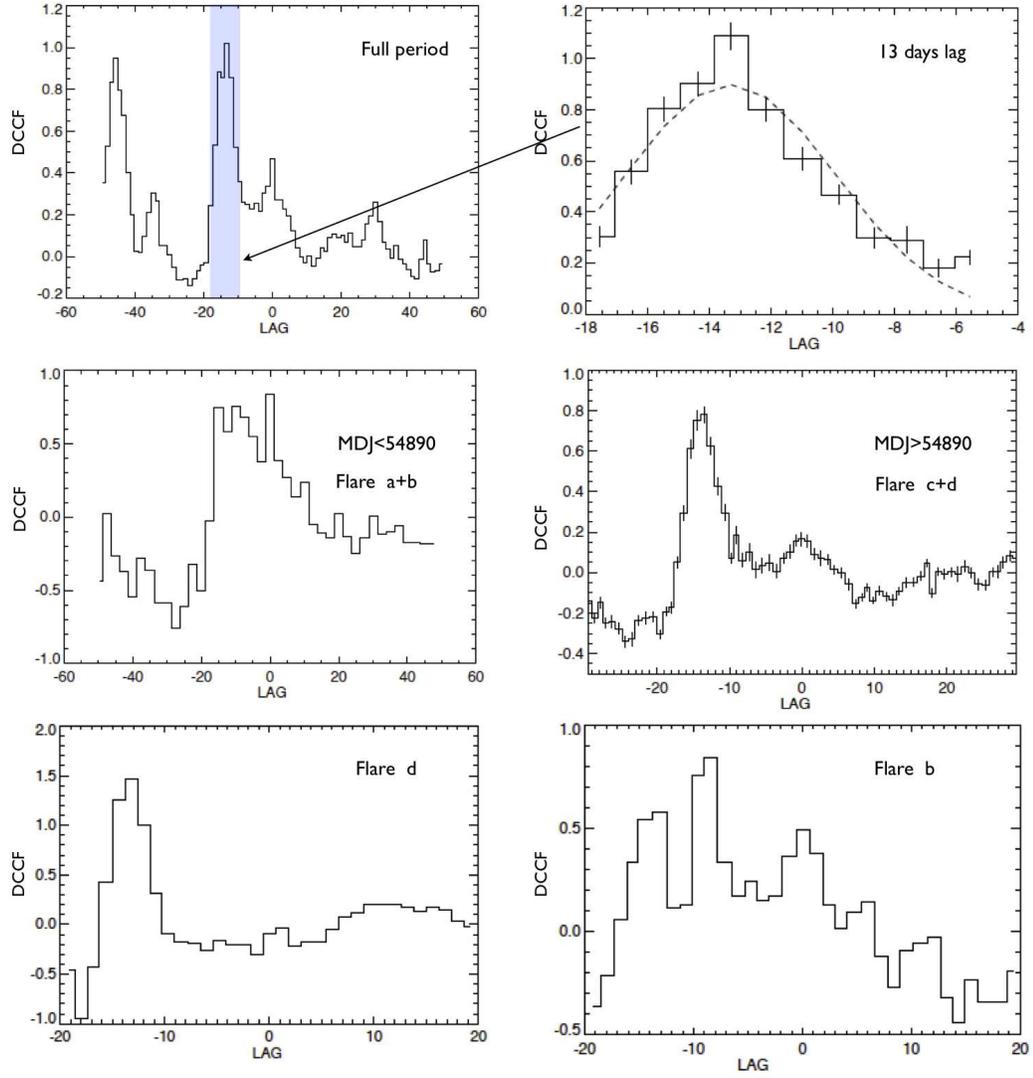}
\end{center}
\caption{
\textit{Top panel, left:} DCCF for the whole analysis period, the shaded box 
concerns the 13 days lag. \textit{Top panel, right:} fit by means of Gaussian 
distribution of the 13 day lag shaded in the left panel, the fit returns a lag 
of -13.4$\pm$ 0.2, gamma leading optical. \textit{Middle panels:} the DCCF for 
MDJ$<$54890 (left panel) and MDJ$>$54890 (right panel). The lag of about 13 days 
with the $\gamma$-ray leading the optical band is still there. \textit{Bottom 
panels:} a lag of about 	13 days (gamma leading optical) is also present for 
flare \textit{d} and \textit{b} individually. } \label{fig:DCCF1} \end{figure} 
\end{minipage}

\begin{minipage}[t]{1\linewidth}
\begin{figure}[H]
\begin{center}
\includegraphics[width=11cm,angle=0]{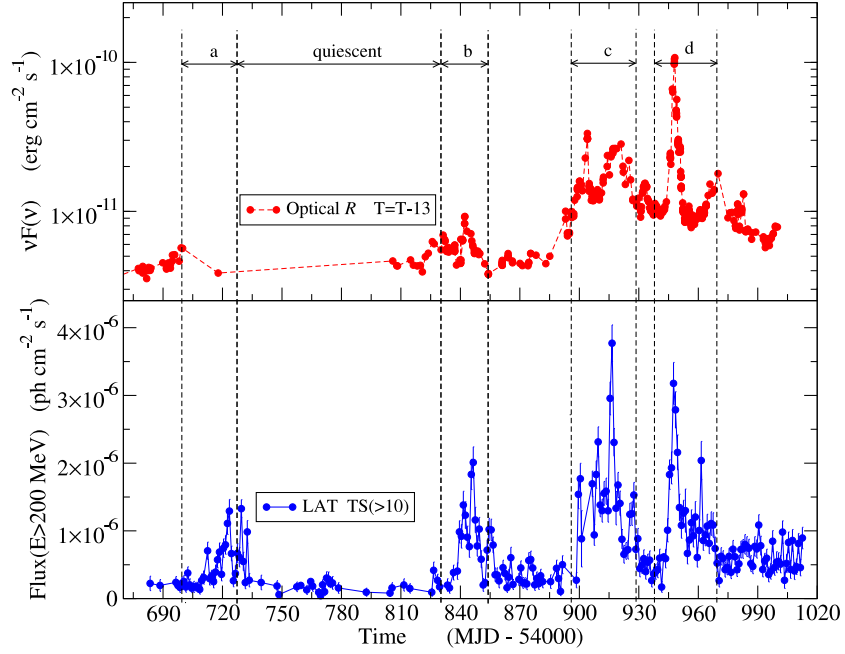}
\end{center}
\caption{{\it Upper Panel:} the optical (\textit{R}) light curve, obtained
shifting the time according to the 13 days \g-ray lag reported in Sect. \ref{sec:Optical and UV results}.
 {\it Lower Panel:} The \textit{Fermi}-LAT light curve }	
\label{fig:LAT-OptR-SHIFTED}
\end{figure}

\begin{figure}[H]
\begin{center}
\epsfig{file=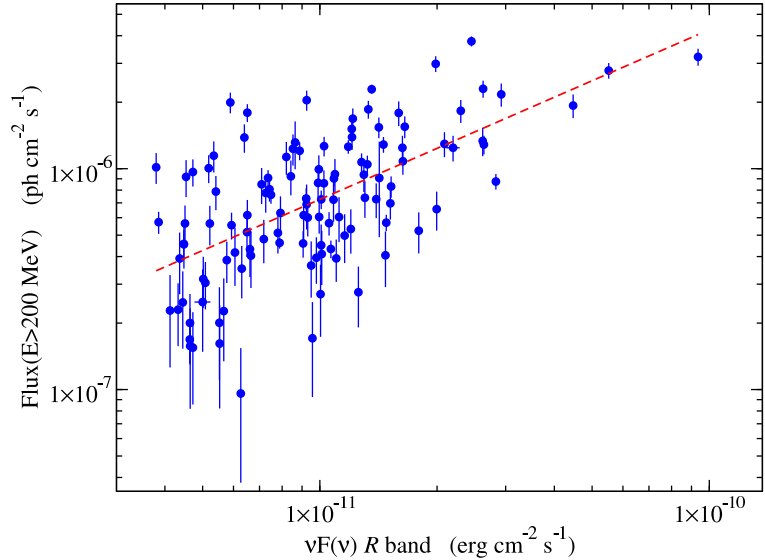,width=10cm,angle=0}
\end{center}
\caption{Scatter plot of the \textit{Fermi}-LAT flux(E$>200$ MeV) vs the $\nu F(\nu)$ in 
the optical $R$ band, using the time-shifted fluxes reported
in Fig \ref{fig:LAT-OptR-SHIFTED} . The dashed line represent the best fit by means of 
power-law model. The correlation coefficient in this case increases
to $r = 0.62$ with a 95$\%$ confidence interval $0.57 \leq r \leq 0.66$.} 
\label{fig:LAT-OptR-CORR-SHIFTED} 
\end{figure} 
\end{minipage}

%

\end{document}